\documentclass[journal,letterpaper]{IEEEtran}

\usepackage{cite}
\usepackage{graphicx}
\usepackage{subfigure}
\usepackage{amsmath} 
\usepackage{multirow}

\begin{document}


\title{Physics-related epistemic uncertainties in proton depth dose simulation}





\author{Maria Grazia Pia, Marcia Begalli, Anton Lechner,
	   Lina Quintieri, and Paolo Saracco 
\thanks{Manuscript received March 30, 2010.}
\thanks{M. G. Pia and P. Saracco are with 
	INFN Sezione di Genova, Via Dodecaneso 33, I-16146 Genova, Italy 
	(phone: +39 010 3536328, fax: +39 010 313358,
	e-mail: MariaGrazia.Pia@ge.infn.it, Paolo.Saracco@ge.infn.it).}
\thanks{M. Begalli is with State University of Rio de Janeiro, Rio de Janeiro,
        Brazil (email: Marcia.Begalli@cern.ch.}
\thanks{A. Lechner is with the Atomic Institute of the Austrian Universities, 
        Vienna University of Technology, Vienna, Austria and CERN, Geneva, 
        Switzerland (email: Anton.Lechner@cern.ch).}
\thanks{L. Quintieri is with INFN Laboratori Nazionali di Frascati, Frascati, 
        Italy (e-mail: Lina.Quintieri@lnf.infn.it).}
}

\maketitle

\begin{abstract}
A set of physics models and parameters pertaining to the simulation of proton
energy deposition in matter are evaluated in the energy range up to
approximately 65 MeV, based on their implementations in the Geant4 toolkit.
The analysis assesses several features of the models 
and the impact of their associated epistemic uncertainties, i.e.
uncertainties due to lack of knowledge, on the simulation results.
Possible systematic effects deriving from uncertainties of this kind are 
highlighted; their relevance in relation to the application environment
and different experimental requirements are discussed, with emphasis
on the simulation of radiotherapy set-ups.
By documenting quantitatively the features of a wide set of simulation
models and the related intrinsic uncertainties affecting the
simulation results, this analysis provides guidance regarding the use
of the concerned simulation tools in experimental applications; it
also provides indications for further experimental
measurements addressing the sources of such uncertainties.


\end{abstract}


\section{Introduction}
\label{sec_intro}
\PARstart{T}{he} 
simulation of the energy deposited by protons in matter is relevant to various
experimental applications; radiotherapeutical applications exploit
its peculiar pattern prior to stopping, exhibiting the characteristic
``Bragg peak'', to deliver a well localized dose to the tumor area \cite{levin}.

Several applications of general purpose Monte Carlo systems, like MCNP
\cite{mcnp,mcnp5,mcnpx}, GEANT 3 \cite{geant3}, Geant4 \cite{g4nim,g4tns},
SHIELD-HIT \cite{shield, shieldhit}, FLUKA \cite{fluka1, fluka2} 
and PHITS \cite{phits}
are documented in the literature concerning this topic, in hadron therapy as
well as in other fields.
A variety of physics options - theoretical models, evaluated data
compilations and values of relevant physical parameters - is available in these
Monte Carlo codes to model the electromagnetic and nuclear 
interactions of protons, and of their secondary products.
While the software implementations 
are specific to each Monte Carlo code, the underlying physics modeling
approaches and data compilations are often common to various
simulation systems.

Some of the physics models and parameters used in the simulation of proton
interactions with matter are affected by epistemic uncertainties
\cite{oberkampf_error}, i.e uncertainties due to lack of knowledge.
They may originate from various sources \cite{oberkampf_predictive}, such as
incomplete understanding of fundamental physics processes, or practical
inability to treat them thoroughly, non-existent or conflicting
experimental data for a physical parameter or model, or the application of a
physics model beyond the experimental conditions in which its validity has been
demonstrated (e.g. at lower or higher energies, or with different target
materials).

The role of epistemic uncertainties in the software verification and validation
process has been the object of research in the context of simulation based on
deterministic methods \cite{oberkampf_error}; these investigations are
motivated by the rigorous risk analysis required by some sensitive applications.
Limited attention has been devoted so far to the role of epistemic uncertainties
in Monte Carlo simulation in particle and nuclear physics, and related
experimental fields.
This paper addresses this topic by illustrating it in a concrete experimental
use case: the simulation of proton depth dose to water for radiotherapy applications.
The simulation configuration involves a realistic model of a therapeutical
proton beam line and beam energies of approximately 65 MeV; this use case is
representative of experimental environments for the treatment of ocular melanoma
\cite{lacassagne,clatterbridge,berlin,catana}.


Due to their intrinsic nature, related to lack of knowledge, epistemic
uncertainties are difficult to quantify \cite{trucano_what}.
Although the characterization of epistemic uncertainty contributions is
needed for many of the issues that feed the reliability model of complex
systems, there is no generally accepted method of measuring epistemic
uncertainties and their contribution to reliability estimations.
A variety of mathematical formalisms \cite{helton_workshop} has been developed
for this purpose; nevertheless, some of the techniques adopted in the context of
deterministic simulations, like interval analysis and applications of
Dempster-Shafer theory of evidence \cite{shafer}, are not always directly
applicable in identical form to the treatment of epistemic uncertainties 
in Monte Carlo simulations.

Sensitivity analysis \cite{trucano_what} is a tool for exploring how different
uncertainties, including epistemic ones, influence the model output
\cite{saltelli}.
This approach is adopted in the study described here: the paper identifies a
set of epistemic uncertainties in physics modeling pertinent to the problem
domain, documents their impact on the simulation results, and discusses their
potentiality to produce systematic effects in relation to the characteristics of
the application environment.


Typically, in statistical analyses epistemic uncertainty is represented as a
set of discrete possible or plausible choices (e.g. model choices) 
\cite{trucano_margins}.
The environment for this kind of study has been realized 
in the context of a Geant4-based application; the characteristic of Geant4 as a
toolkit, encompassing a wide variety of physics models, along with the feature
of polymorphism characterizing the object oriented programming paradigm, allow
the configuration of the simulation with a large number of different physics
options in the same software environment. This versatility makes Geant4 a
convenient playground to evaluate the effects of a number of physics
alternatives on the experimental use case under study.
The sensitivity analysis documented in this paper, which examines the response
of the system to a wide set of modeling approaches, plays a conceptually similar
role to the interval analysis method applied in deterministic simulation, where
parameters subject to epistemic uncertainties are varied within bounds.
As mentioned in the previous paragraph, sensitivity analysis as
applied to the context described in the following sections contributes
to identify and quantify possible systematic effects in the simulation; 
it cannot infer anything about the validity of any of the physics 
models, for which experimental data would be needed.

The physics modeling options available in Geant4 for the use case under study
are broadly representative of the body of knowledge in the problem domain; the
analysis described in this paper, albeit performed in the context of a specific
Monte Carlo system, provides elements of interest for other simulation
environments as well.


Epistemic uncertainties are in principle reducible \cite{oberkampf_error};
the analysis documented in this paper identifies areas where experimental
measurements could reduce them, and highlights their requirements of accuracy
and other features to improve the knowledge embedded in the current
simulation models.

Preliminary assessments relevant to this study were reported in
\cite{bragg_nss2008}.

\section{Geant4-based simulation of proton depth-dose profiles}

The Geant4 toolkit includes various physics modelling options relevant to the
application domain considered in this paper; they concern stopping powers,
multiple scattering, cross sections and final state models of elastic and
inelastic nuclear interactions of the primary protons, as well as a variety of
electromagnetic and hadronic models for the secondary particles resulting from
proton interactions with matter.

Several Geant4-based simulations of proton therapy set-ups, like
\cite{paganetti2003}-\cite{paganetti2008}, have documented satisfactory
agreement with experimental depth dose measurements in various configurations of
beam lines and detectors.
Some open issues remain, which are generally shared by simulations based on
other Monte Carlo systems as well, due to common physics modeling approaches in
the codes and similar practices in the experimental domain.

There is evidence in the literature of different features produced by Geant4
physics models in the energy range below 100 MeV
\cite{emnist,chin,quesada,ivanchenko2,apocalor2008}; the discrimination of their
accuracy is made difficult by the still incomplete software validation, which is
often hindered by the limited availability of experimental data, or their
controversial characteristics.
The same problem affects other Monte Carlo codes as well.
Furthermore, some shortcomings are present in the comparisons of proton
therapy simulations with experimental data.

The physics configuration used in the simulation and the Monte Carlo kernel
version are either undocumented, or incompletely documented, in a number of references;
therefore it is not always possible to relate the results documented in the
literature to the physics options and software implementations which produced
them, thus hindering the reproducibility of the results.
The comparison between simulated and experimental data is limited to qualitative
appraisal in most cases; the lack of statistical analysis in many articles
prevents the reader from appraising the significance of the compatibility
between simulation and experimental data, and the relative merits of the
simulation models in comparison to measurements taken in different experimental
configurations and characterized by different uncertainties.

It is a feature of proton therapy simulations that the
beam parameters, namely the beam energy and energy spread, are not known in
typical experimental set-ups of this domain with sufficient accuracy to base the
simulation on their nominal values \cite{paganetti2004}.
The precise knowledge of the beam energy is not critical to clinical
applications; in that context
what is of interest is the proton range, which can be measured very accurately.
In common practice, the beam parameters input to the simulation are adjusted
so that the simulation results best fit the measured depth dose profile
\cite{paganetti2003,paganetti2004,pablo,newhauser,shipley,gudowska}; this
procedure affects the significance of further comparisons between simulated and
experimental data.
Experimental techniques have been developed to measure the energy of a proton
beam from radiotherapy accelerators with greater precision \cite{brooks}, but
they are not commonly exploited in connection with simulation validation
studies.

The comparison of simulated and experimental Bragg peak profiles is also
sensitive to the normalization procedures, which are often applied in
experimental practice to relate simulated and measured data; this topic is
discussed in detail in \cite{bragg_norma}.
 
These shortcomings do not severely affect current clinical practice, where
Monte Carlo simulation plays a role of verification and optimization.
More demanding requirements about the predictive capabilities of the simulation
may derive from new perspectives, such as the use of Monte Carlo simulation in
treatment planning, which is the object of ongoing research
\cite{tps_infn,mctps, tilly_worth, webb}, or other disciplines, like radiation
protection.

Epistemic uncertainties undermine the predictive role of simulation software; by
identifying and quantifying them, the analysis described in the following
sections is propaedeutic to further experimental and theoretical efforts to
reduce them, or at least to control their effects.

\section{Overview of relevant physics models}
\label{sec_g4models}

A brief overview of the Geant4 physics models relevant to the use case addressed
in this paper is given below to facilitate the understanding of the results
presented in the following sections.
In the context of Geant4, particle interactions with matter are represented by
processes, which can be implemented through different models \cite{g4nim}.
Similarities with the modeling approaches in other Monte Carlo codes
are discussed; they show that the body of knowledge, as well as knowledge gaps, 
are largely shared across a variety of simulation systems.
The information summarized in this section is necessarily succinct;
further details can be found in the cited references.


\subsection{Electromagnetic interactions}
\label{g4em}
Geant4 electromagnetic package \cite{em_mc2000} includes two packages
relevant to the experimental use case considered: \emph{Standard} and
\emph{Low Energy}.

The proton ionisation process in the \emph{Low Energy} electromagnetic package
\cite{lowe_chep,lowe_nss} is articulated through models \cite{lowe_p} based on
the free electron gas model \cite{lindhard} at lower energy ($<$1 keV), on
parameterisations at intermediate energy, and on the Bethe-Bloch equation at
higher energy.
Alternative parameterisation models, identified in the following as
\emph{ICRU49}, \emph{Ziegler77}, \emph{Ziegler85} and \emph{Ziegler2000}
implement electronic and nuclear stopping powers based on \cite{icru49,
ziegler77, ziegler85, srim2000}; their different energy ranges of applicability
are documented in the respective references.
The configuration of the hadron ionisation process is identified in the
following sections through the selected parameterisation model.
The process also deals with the emission of $\delta$-rays.

The Geant4 \emph{Low Energy} electromagnetic package includes processes
\cite{lowe_e,lowe_ion} to handle the secondary particles resulting from proton
interactions: electrons, photons and ions.
Models for electrons and photons are based on data libraries (EEDL \cite{eedl}
and EPDL97 \cite{epdl97}) and on analytical formulations originally developed
for the Penelope \cite{penelope} Monte Carlo code; both options account for the
atomic relaxation \cite{relax} following the primary processes.
Models based on the parameterisations in \cite{ziegler77}, \cite{ziegler85}, and
on \cite{icru73} are available for ions.

The main features of the Geant4 \emph{Standard} electromagnetic package are
documented in \cite{standard}.
More recently a model \cite{radphyschem} for proton ionisation based
on \cite{icru49} was implemented in this package; this physical
approach is the same as the one already present in the \emph{Low Energy} 
package.
The handling of energy loss fluctuations is implemented in this package; it is
based on the model adopted in GEANT 3 \cite{lassila} and was updated in Geant4
8.3 \cite{standard_chep07}.

An assessment of Geant4 electromagnetic processes against the NIST (United
States National Institute of Standards) reference data can be found in
\cite{emnist}.
The validation of Geant4 low energy electromagnetic processes against precision
measurements of electron energy deposition is documented in \cite{sandia_tns}.

The \emph{Standard} package includes implementations \cite{urban2002,urban2006}
of the multiple scattering process.
In the early Geant4 versions a generic process (\textit{G4MultipleScattering}),
based on the Lewis \cite{lewis} theory, was applicable to any charged particles;
it has been complemented by a process specialized for hadrons
(\textit{G4hMultipleScattering}) in Geant4 8.2, and one
specific to electrons (\textit{G4eMultipleScattering}) in Geant4 9.3.
The specialized multiple scattering processes are intended to replace
the generic one in future Geant4 releases \cite{relnotes93}.
These processes can be configured with various models (e.g.
\textit{G4UrbanMscModel90}, \textit{G4UrbanMscModel92} and
\textit{G4UrbanMscModel93}), which involve some empirical parameters
\cite{elles}.
Early implementations of Geant4 multiple scattering were validated against
experimental muon data
at low \cite{xmm,muscat} and high \cite{pedro_l3} energy;
\cite{pedro_l3} showed better accuracy of Geant4 multiple scattering model (as
implemented in Geant4 1.0) with respect to the GEANT 3 model based on Moli\`ere
theory.
Recent studies, like \cite{sandia_tns,cmscalo}, have highlighted issues in the
evolution of energy deposition patterns involving electron-photon interactions
simulated with different Geant4 versions;
the observed effects have been ascribed to modifications to Geant4 multiple
scattering algorithm.
The literature concerning recent evolutions of Geant4 multiple scattering is
focused on issues related to the dependence of the simulation on 
transport step size
and parameters of the algorithm \cite{radphyschem,elles,standard_chep07}.
Io the best of the authors' knowledge,
the validation of Geant4 proton multiple scattering is not documented in
literature yet for the energy range relevant to this study.


The Geant4 \emph{Standard} and \emph{Low Energy} electromagnetic packages 
are concerned by the same epistemic uncertainties affecting electromagnetic 
physics modeling relevant to the use case studied in this paper, which are 
analyzed in sections \ref{sec_ionipot} and \ref{sec_stopping}.
Differences in the outcome of simulations using the two packages may derive from 
features specific to each package, which are not associated with epistemic 
uncertainties in 
the physics domain pertinent to the use case examined in this paper;
they could be due to numerical features, like the number of bins
in the look-up tables used in the simulation and their interpolation, or algorithms 
and modelling choices specific to each package.
A complete documentation and analysis of different features of Geant4 
\emph{Standard} and \emph{Low Energy} electromagnetic packages is outside the
scope of this paper.

Other Monte Carlo codes used for hadron therapy simulation adopt
similar approaches for stopping power calculation at high and low
energies.
At intermediate energies, stopping powers based on ICRU 49 Report are
implemented in SHIELD-HIT; an improvement to include them in FLUKA is documented
in \cite{parodi}, but it does not appear to be released yet in the current
version of FLUKA (FLUKA-2008) at the time of writing this paper. PHITS handles
proton ionization according to the SPAR code \cite{spar}, while MCNP uses an
energy weighted average between the high and low energy calculations
\cite{mcnp5_stop}, which adopt the same methodology as in SPAR.
PHITS \cite{nose}, SHIELD-HIT \cite{gudowska} and MCNPX
do not model $\delta$-ray emission.

GEANT development was frozen with the 3.21 version in 1994; the code is no
longer supported, but it is still used for hadron therapy 
developments \cite{faiza}.
GEANT 3 simulated proton energy loss based on the Bethe-Bloch equation
and dealt with $\delta$-ray production from ionization.

MCNP multiple Coulomb scattering treats soft and hard interactions separately
\cite{mcnp_rad}: soft collisions are described using a continuous scattering
approximation; a small number of hard collisions are simulated directly.
Multiple scattering is based on Moli\`ere theory in FLUKA and PHITS
\cite{phits2006}; details of the algorithm in FLUKA are described in
\cite{fluka_mscatt}.
SHIELD-HIT simulates multiple scattering on the basis of a gaussian model
\cite{gudowska}, which gives the correlated value of the angular deviation and
lateral displacement of the scattered particle.
GEANT 3 provided two options for multiple scattering simulation, respectively
based on a Gaussian approximation and on Moli\`ere theory \cite{geant3}.



\subsection{Hadronic interactions}
\label{sec_g4hadronic}

The Geant4 hadronic package addresses the complexity of nuclear interactions
through a software framework \cite{wellisch_frameworks}.
The baseline design can accommodate multiple implementations of cross sections
and final state models associated with a process, which are either complementary
in their energy range coverage or alternative in their modelling approach;
processes and models are meant to be handled polymorphically through their
respective base classes.

\subsubsection{Elastic scattering}
\label{sec_g4elastic}

Geant4 includes various elastic scattering processes:
\emph{G4HadronElasticProcess}, \emph{G4UHadronElasticProcess},
\emph{G4QElastic} and \emph{G4WHadronElasticProcess}; the latter
was released in Geant4 9.3 with the purpose \cite{relnotes93} of
allowing models for elastic scattering to be treated in a similar way to
inelastic models.
A \emph{G4DiffuseElastic} \cite{diffuse} process was also released in Geant4 9.3;
the energy range of its applicability is not explicitly specified in \cite{diffuse},
but, since this process appears applied at energies of 1 GeV and above in the 
associated reference, one is led to assume that it is pertinent to higher energies
than those relevant to the use case studied in this paper.
This inference manifests an epistemic uncertainty in the applicability domain
of this process.

The \emph{G4HadronElasticProcess} class of the hadronic \emph{processes}
package handles cross section and final state calculation according to the 
software design of \cite{wellisch_frameworks}.

It can be configured with the \emph{G4HadronElasticDataSet} class, derived from
\emph{G4VCrossSectionDataSet} and included in the Geant4 hadronic
\emph{cross\_sections} package, which implements total elastic scattering cross
sections derived from GHEISHA \cite{gheisha}.

The scattering can be configured through several models, such as
\emph{G4LElastic}, \emph{G4ElasticCascadeInterface} and \emph{G4HadronElastic}.
\emph{G4LElastic}, included in the hadronic \emph{models} package, is based on
GHEISHA algorithms reengineered in Geant4; it is not meant to conserve energy
and momentum on an event-by-event basis, but only on average.
\emph{G4ElasticCascadeInterface}, identified in the following as ``Bertini
elastic'', is included in the \textit{cascade} package of the \textit{cascade}
package of hadronic \textit{models}; it derives from
\emph{G4IntraNuclearTransportModel} and implements a model based on the INUCL
\cite{inucl} code.
The \emph{G4CascadeElasticInterface} class in the same package activates both
elastic and inelastic interactions.
\emph{G4HadronElastic} \cite{g4uhadronelastic}, included in the
\emph{coherent\_elastic} package of hadronic \textit{models}, combines elements
originally developed for CHIPS (Chiral Invariant Phase Space)
\cite{chips1,chips2} with other modelling approaches; it aggregates a
\emph{G4VQCrossSection} object belonging to CHIPS and a
\emph{G4ElasticHadrNucleusHE} model.

The \emph{G4QElastic} process, known as ``CHIPS elastic'',  delegates
the calculation of cross sections to the \emph{G4QElasticCrossSection} class,
derived from \emph{G4VQCrossSection}, and implements its own scattering
algorithm.
All the related classes are included in the \textit{chiral\_inv\_phase\_space} 
package of hadronic \textit{models}.

The \emph{G4UHadronElasticProcess} process \cite{g4uhadronelastic}, included in the
\emph{coherent\_elastic} package of hadronic \textit{models}, is
meant to be configured with the dedicated \emph{G4HadronElastic} model;
this configuration is referred to in the following as ``U-elastic''.

The limited documentation in the literature of Geant4 elastic scattering models
and other codes does not facilitate the appreciation of their characteristics,
nor the identification of the experimental data with respect to which
some of the simulation models, especially those implementing parameterisations, 
may have been calibrated.
Although improvements to Geant4 elastic scattering modeling are mentioned in
\cite{apocalor2008}, hardly any validation of the available models is documented
in the literature in the energy range relevant to the use case under study.
The use of these models in the simulation is a source of epistemic
uncertainty, due to the lack of knowledge of their accuracy in the
energy range pertinent to this study. 


\subsubsection{Non-elastic interactions}

Inelastic hadron-nucleus scattering is handled in Geant4 through processes
specific to each particle type.
Processes for protons, neutrons, deuteron, triton and $\alpha$ particles
are relevant to this study.

Total inelastic cross sections derived from GHEISHA \cite{gheisha} are available 
in Geant4 through \emph{G4HadronInelasticDataSet} for all hadrons, $\alpha$
particles, deuteron and triton; alternative implementations based on
\cite{axen}, \cite{barashenkov} are available for some energy ranges
and target materials.
Cross sections describing neutron-nucleus scattering with higher precision below
20~MeV are available in \emph{G4NeutronHPInelasticData}. Specialized cross
sections, based on \cite{kox}-\cite{tripathi4}, are available for ions.


Parameterised and theory-driven \cite{amelin} models of nuclear
inelastic scattering are available in Geant4 for protons and other particles,
concerning the energy range pertinent to this study.

Geant4 Low Energy Parameterised models (LEP), originating from GHEISHA
\cite{gheisha}, handle protons, neutrons, pions, $\alpha$ particles,
deuterons and tritons.

The CHIPS \emph{G4QInelastic} inelastic process \cite{chips1,chips2} implemented
in Geant4 is applicable to hadron inelastic scattering in the energy range
pertinent to this study.

Various options of theory-driven models describe the phases of intranuclear 
transport, preequilibrium and nuclear deexcitation in Geant4.
Other Monte Carlo codes used in proton therapy applications use a similar 
multi-stage approach to simulate proton inelastic interactions.
The primary proton energy in this study lies at the edge of what is commonly
considered as the range of transition between intranuclear cascade and
preequilibrium models.

Geant4 includes three cascade models, each one with further associated models
describing the lower energy phases: they are known as Binary
\cite{binary}, Bertini \cite{bertini1,bertini2} and Li\`ege \cite{liege} cascade
models.

The Binary cascade model \cite{binary} adopts a hybrid approach between a
classical cascade code and a quantum molecular dynamics model.
It handles the preequilibrium phase through the Precompound model
(\emph{G4PreCompoundModel}) \cite{lara}, whose implementation is based on
Griffin's exciton model \cite{griffin,griffin2}; this model can be activated in
a simulation application either through the Binary cascade model or as an
independent model.
The nuclear deexcitation associated with the Precompound model can be configured
with various options: they include an evaporation model based on
Weisskopf-Ewing's \cite{weisskopf,weisskopf-ewing} theory, exploiting
Dostrovsky's computational approach \cite{dostrovsky}, the Generalized
Evaporation Model (GEM) \cite{gem} (also used by PHITS), and the
optional activation of the Fermi break-up \cite{lara}.

A similar approach is adopted in SHIELD-HIT, whose default model considers a
fast cascade stage, which brings the interaction between the projectile and
target to a sequence of binary collisions \cite{toneev_gudima}; this stage is
followed by preequilibrium \cite{gudima} and deexcitation of residual nuclei,
with Fermi break up of light nuclei and evaporation.

The Geant4 Bertini Cascade implementation \cite{bertini1,bertini2} is a
reengineering of the INUCL code \cite{inucl}, which is based on Bertini's
approach \cite{bertini} to intranuclear transport; it handles the preequilibrium
phase based on Griffin's exciton model and the evaporation phase based on
Weisskopf's and Dostrovsky's approach.
The preequilibrium part of INUCL is based on the Cascade Exciton Model (CEM)
\cite{cem}, which is one of the options of MCNPX for proton transport and is
also implemented in SHIELD (as well as in other codes) \cite{mashnik_1998}.

A cascade model based on Bertini's scheme, derived from the HETC \cite{hetc}
implementation in LAHET \cite{lahet}, is available in MCNPX to handle protons,
besides the ISABEL \cite{isabel1,isabel2} and CEM options.
The HETC model was also interfaced to GEANT 3 through CALOR \cite{calor}.

The INCL Intranuclear Cascade \cite{liege} model, better known as Li\`ege
Cascade, has been reengineered in Geant4 together with the associated ABLA
evaporation model \cite{abla}; it was first released in Geant4 version 9.1
\cite{inclabla} and further improved in Geant4 9.3.
INCL is included \cite{cugnon} in the LAHET \cite{lahet} code system used by MCNPX.
Although, according to \cite{inclabla}, INCL is meant for energies above 200 MeV, 
satisfactory applications at energies of a few tens MeV are reported in the literature 
\cite{boudard1,boudard2,boudard3}.

The simulation of low energy proton interactions in PHITS is based \cite{sato} on the
MCNP4C and NMTC \cite{nmtc} codes; the latter incorporates Bertini's cascade
model \cite{bertini} for nucleon and meson transport.

FLUKA handles inelastic scattering through PEANUT \cite{peanut} in the energy
range relevant to the use case under study; it involves a sequence of
intranuclear cascade followed by preequilibrium and deexcitation.
The preequilibrium model is based on the formalism developed by Blann
\cite{blann} with some modifications \cite{ferrari_sala_ts1998}; the evaporation
model is based on Weisskopf-Ewing's approach \cite{ballarini2005}; Fermi break-up 
is modeled for light nuclei \cite{ballarini2005}.

Hadronic interactions were not handled by GEANT 3; their treatment was
delegated to external packages (GHEISHA, CALOR and an early version of 
FLUKA) interfaced to it.


Limited documentation regarding the validation of Geant4 inelastic scattering
models relevant to the use case of this study is available in the
literature.
Some comparisons with experimental data are reported in
\cite{quesada}, \cite{ivanchenko2}, \cite{apocalor2008},
\cite{g4uhadronelastic},\cite{binary}, \cite{koi2006}, \cite{wellisch_chep2003}: 
although most of the results shown in these references are not directly related to
the use case investigated in this paper,
they demonstrate the ongoing validation efforts in this domain.

\begin{table*}
\begin{center}
\caption{Proton physics modelling options in the simulation}
\label{tab_produced}
\begin{tabular}{|l|l|l|l|}
\hline 
{\bf Physics domain} 	& \textbf{Option} 	& \textbf{Process class}  	& \textbf{Model class} 	\\
\hline
Proton 		& ICRU49			& G4hLowEnergyIonisation	& G4hICRU49p		\\
stopping		& Ziegler77			& G4hLowEnergyIonisation	& G4hZiegler1977p	\\
powers		& Ziegler85			& G4hLowEnergyIonisation	& G4hZiegler1985p	\\
			& Ziegler2000		& G4hLowEnergyIonisation	& G4hSRIM2000p		\\
\hline
Multiple scattering	& Generic multiple scattering & G4MultipleScattering    	& G4UrbanMscModel92	\\
			& Hadron multiple scattering  & G4hMultipleScattering   & G4UrbanMscModel90	\\
\hline
Hadronic 		& LEP				& G4HadronElasticProcess	& G4LElastic		\\
elastic		& U-elastic			& G4UHadronElasticProcess	& G4HadronElastic	\\
scattering		& Bertini-elastic		& G4HadronElasticProcess	& G4ElasticCascadeInterface \\
			& CHIPS-elastic		& G4QElastic			& 			\\
\hline
Hadronic inelastic	& Default 			& G4ProtonInelasticProcess	& G4HadronInelasticDataSet \\
cross sections	& Wellisch \cite{axen}	  	& G4ProtonInelasticProcess	& G4ProtonInelasticCrossSection	\\
\hline
Hadronic 		& LEP				& G4ProtonInelasticProcess	& G4LEProtonInelastic	\\
inelastic		& Precompound		& G4ProtonInelasticProcess	& G4PreCompoundModel	\\
scattering 		& Precompound-GEM	& G4ProtonInelasticProcess	& G4PreCompoundModel, G4ExcitationHandler \\
    		         & Precompound-Fermi break-up & G4ProtonInelasticProcess	& G4PreCompoundModel, G4ExcitationHandler \\
 		         & Binary cascade   	& G4ProtonInelasticProcess	& G4BinaryCascade \\
			& Bertini cascade		& G4ProtonInelasticProcess	& G4CascadeInterface	\\
			& Li\`ege			& G4ProtonInelasticProcess	& G4InclAblaCascadeInterface	\\
			& CHIPS-inelastic		& G4QInelastic			& 			\\
\hline
\end{tabular}
\end{center}
\end{table*}

\subsection{Epistemic uncertainties in the simulation models}

Epistemic uncertainties in the physics simulation models arise from various
sources.

In some cases, the lack of knowledge concerns the value of a physical parameter:
this is the case, for instance, for the mean water ionization potential, for
which various values, originating from experimental measurements or theoretical
calculations, are documented in the literature.

Other sources of uncertainty are associated with values, used in the simulation,
deriving from parameterisations, or fits to experimental data (which may be
inspired by theoretical motivations): in the present study this concerns, for
instance, the cross sections of nuclear elastic and inelastic interactions, and
proton stopping powers.
In these cases the uncertainties derive from the measurements themselves, the
criteria by which the data are selected for the fit, and from the
parameterisation or fitting process.

Some models may embed parameters or, more generally, features, which are
adjusted in the software implementation according to empirical procedures: from
calibration with respect to experimental data to educated guesses, in the
absence of pertinent measurements.
This is the case, for instance, for the Geant4 multiple scattering model and forsome
hadronic interaction models.
In this respect, and also for models based on parameterisations or
fits to experimental data, an important issue is the distinction between the
calibration and validation of simulation models; for the reader's convenience,
these concepts pertaining to simulation epistemology are briefly recalled here.
Calibration is the process of improving the agreement of a code calculation
with respect to a chosen set of benchmarks through the adjustment of parameters
implemented in the code \cite{trucano_what}; in the Monte Carlo simulation
jargon this process is also known as ``tuning''.
Software validation is defined in the IEEE Standard for Software Verification
and Validation \cite{ieee_vv}.
This generic definition is adapted to specific application domains with some
slight variants; regarding simulation, validation is usually
intended as the process of determining the degree to which a model is an
accurate representation of the real world from the perspective of its intended
uses, or of confirming that the predictions of a code adequately represent
measured physical phenomena \cite{trucano_what}, \cite{asc}.

The limited documentation in the literature of the calibration of the physics
models implemented in Monte Carlo codes does not facilitate the understanding
whether some of the comparisons with experimental data reported in the
literature document calibration results and their experimental benchmarks, or
model validation.

The use of a simulation code for predictive purposes outside the scope of its
validation necessitates extrapolation beyond the understanding gained strictly
from experimental validation data.
This type of uncertainty in our inference is primarily epistemic.

Regarding the energy range of nuclear interactions relevant to the use case
considered in this paper, a long debate has been going on for decades in the
literature about different theoretical preequilibrium models, respectively
based on the so-called ``exciton" and ``hybrid" approaches
\cite{bisplinghoff}.
These discussions involve subtle theoretical arguments;
however, it has been acknowledged that, whichever theoretical approach
is chosen to model equilibrium emission, the effects of overly
simple or untested assumptions can be compensated by means of other
uncorrelated phenomenological parameterisation of the model.
This ongoing theoretical debate is not expected to be relevant to the use case
addressed in this study, since the differences of the various theoretical 
approaches are expected to affect mainly exclusive channels \cite{koning}, 
with negligible effects on the resulting deposited energy, which is the object
of this paper.
The sources of epistemic uncertainties in the simulation reside
in the phenomenological features of the nuclear model implementations, rather
than in the choice of theoretical approach to preequilibrium modeling.

The analysis described in the following sections identifies sources of 
epistemic uncertainties in the physics domain of the simulation
and evaluates systematic effects on the simulation outcome
associated with them.

\section{Software configuration}

\subsection{Simulation}
\label{sec_simu}

The simulation application includes components responsible for the configuration
of the geometry and materials of the experimental set-up, the generation of the
proton beam, the selection of the physics features to be used in the particle
transport, the collection of relevant observables in dedicated objects, and the
control of the user's interaction with Geant4 kernel at various stages of the
execution.

The geometry configuration encompasses a realistic model of a proton therapy
beam line and a volume, placed at the exit of it, where the energy deposited by
the proton beam is scored.
The beam line model exploits the code of the geometry and material composition
of a real-life proton therapy facility \cite{catana}, which is publicly released
in the Geant4 \emph{hadrontherapy} \cite{pablo} example; the implementation of
the beam line geometry as in Geant4 8.1 was used for all the simulation
productions.
The scoring volume consists of 4 cm cube, filled with water; its size
is adequate to contain the formation of the Bragg peak of energy loss produced
by the proton beam.
This volume is defined ``sensitive'' \cite{g4nim} in Geant4 terms.
A readout geometry, longitudinally segmented in 200~$\mu m$ thick slices, is
superimposed to the mass geometry of the sensitive volume; the longitudinal
segmentation determines the resolution of the simulation in the location of the
Bragg peak, and mimics the resolution of typical measurements
with ionization chambers in experimental practice.
The energy deposit profile is scored through Geant4 hits objects.
The figures of longitudinal energy deposition profiles included in 
the following sections show the energy deposited in each slice of the 
readout geometry by primary protons and secondary particles, integrated over
all the events in the simulation run.



Primary protons are generated according to a Gaussian distribution with 63.95
MeV mean energy and 300~keV standard deviation, unless differently specified in the following
sections.

A variety of physics options is configured by means of a software design
exploiting Geant4 \emph{G4VModularPhysicsList} and \emph{G4VPhysicsConstructor}
as base classes.
A class derived from Geant4 kernel's \emph{G4VModularPhysicsList} is
responsible for driving the physics selection in the simulation application: it
assembles a physics configuration by means of subclasses of
\emph{G4VPhysicsConstructor}, each one responsible for instantiating the physics
processes and models pertinent to a particle type (or a group of particles, like
ions) and an interaction type (electromagnetic, hadronic elastic or hadronic
inelastic).

The proton physics options and corresponding Geant4 classes
evaluated in this study are summarized in Table \ref{tab_produced}.

For convenience, a subset of \emph{G4VPhysicsConstructor} subclasses,
corresponding to the selection in Table \ref{tab_refconf}, has been defined
as a reference configuration for this study.

\begin{table}
\begin{center}
\caption{Reference physics configuration in the simulation}
\label{tab_refconf}
\begin{tabular}{|l|c|c|c|c|}
\hline 
{\bf Particles} & \textbf{Physics selection} 		\\
\hline
Electrons	& Low energy electromagnetic package, 	\\
Photons	& library-based models			\\
\hline
Positrons	& Standard electromagnetic package			\\
\hline
Protons	& \textit{G4hLowEnergyIonization}, \textit{ICRU49} parameterisations \\
		& \textit{G4UHadronElastic} process, \textit{G4HadronElastic} model \\
		& \textit{G4ProtonInelasticProcess}, Precompound model \\
               	& Inelastic cross sections as in \cite{axen}	\\

\hline
Neutrons	& \textit{G4UHadronElastic} process, \textit{G4HadronElastic} model \\
		& \textit{G4NeutronInelasticProcess}, Precompound model \\
		& \textit{G4HadronCaptureProcess}, LEP model 	\\
		& \textit{G4HadronFissionProcess}, LEP model	\\
 		& Inelastic cross sections as in \cite{axen}	\\
\hline
Deuteron	& \textit{G4hLowEnergyIonization}, \textit{ICRU49} parameterisations (scaled) \\
Triton	         & \textit{G4UHadronElastic} process, \textit{G4HadronElastic} model \\
$\alpha$	& Inelastic process specific to each particle, LEP models\\
		& Tripathi and Shen cross sections		\\
\hline
Charged  	& \textit{G4MultipleScattering} process \\
\hline
\end{tabular}
\end{center}
\end{table}

In all the productions the interactions of secondary particles are simulated as
in Table \ref{tab_refconf}, unless differently specified in the following
sections.

Options regarding the water ionization potential can be introduced in
the simulation by setting the value of this parameter through 
the public interface of the \textit{G4Material} class.


The secondary particle production threshold is set at 1 $\mu$m (in range) in
the water volume and 10 $\mu$m in the passive components of the
experimental set-up.  
The results presented in the following sections are generated 
with 10 $\mu$m maximum step size; the step limit is set approximately an order
of magnitude smaller than the thickness of the slices in the readout geometry to
ensure adequate precision in scoring the longitudinal energy deposition profile.

\subsection{Simulation production}
\label{sec_production}


Simulated data were produced by instantiating various combinations of
physics constructor objects in the simulation application corresponding to
the options listed in Table \ref{tab_produced}.

The simulation production concerned a set of representative physics
configurations, each differing from the reference one of Table
\ref{tab_refconf} by one modeling feature only; this strategy allowed the
unambigous attribution of the differences observed in each simulation to the
physics feature specifically under investigation.
One million events were generated for each physics configuration. 

Data samples corresponding to all the options listed in Table \ref{tab_produced}
were produced with Geant4 9.3, the latest version available at the time
of writing this paper.
Data samples corresponding to a subset of physics configurations were also
produced with three other Geant4 versions: 8.1p02, 9.1 and 9.2p03.
Versions 8.1p02 and 9.1 were involved in the validation of electron energy
deposition in \cite{sandia_tns}; Geant4 8.1 was previously used for the assessement
of some Geant4 physics models in \cite{paganetti2008} (subject to the same
beam settings adjustments as in \cite{paganetti2004}), while
Geant4 9.1 is still widely used in production mode by various experiments.
Geant4 9.2p03 is an updated version of
Geant4 9.2 including corrections; it was released two months later than Geant4 9.3.

The simulations were run on Intel Core2 Duo Processors E6420, equipped with
2.13~GHz CPU and 4~GB RAM, Scientific Linux 4 operating system and gcc
3.4.6. compiler.




\subsection{Data analysis}
\label{sec_analysis}

Data analysis objects were handled in the simulation application code through
AIDA \cite{aida} abstract interfaces; the iAIDA \cite{iAIDA}
implementation was used.

The results of the various simulation configurations were compared to the
outcome of the reference one.
No normalization was performed on the Bragg peak profiles to be compared (unless
explicity stated in the following sections): therefore the distributions, which
originate from the same number of primary protons in all the simulation
configurations, allow the appreciation not only of differences in shape and peak
location, but also of absolute effects, like the proton acceptance in the
sensitive volume and the total energy deposited in it.

Within a given Geant4 version, any observed difference in the results is to be
ascribed to intrinsic effects of the physics models activated in the simulation.
To the best of the authors' knowledge, the Geant4 transport kernel did not
modify its behavior through the versions considered in this study; differences
observed in comparisons across different Geant4 versions reflect the evolution
of the physics model implementations.

The differences associated with the various simulation configurations were
quantitatively estimated by means of statistical methods; the comparisons exploited 
non-parametric, unbinned goodness-of-fit tests available in the
Statistical Toolkit \cite{gof1,gof2}.
Different algorithms were applied to each test case to evaluate possible
systematic effects on the results due to particular features of the tests: the
Anderson-Darling \cite{anderson1952,anderson1954}, Cramer-von Mises
\cite{cramer1928,vonmises1931,fisz1960} and Kolmogorov-Smirnov
\cite{kolmogorov1933,smirnov1939} tests.
The null hypothesis in the test process assumed the equivalence between the
distributions subject to comparison.
The critical value for the rejection of the null hypothesis was set at 0.05,
unless differently specified; in the following sections the expression ``95\%
confidence level'' is used to indicate 0.05 significance of the test.


Goodness-of-fit tests compared the whole longitudinal distributions of energy
deposition, as well as the distributions corresponding to the left and right
branches of the longitudinal profiles, i.e. at penetration depths respectively
up to and beyond the peak position, to ascertain the compatibility of the
data samples in detail.

The comparison results mentioned in the following sections concern the left
branch of the energy deposition profiles, unless differently specified.
Due to the mathematical features of the test statistics and empirical
distribution functions, the left branch of the profiles provides the most sensitive
test case to highlight differences in the distributions subject to comparison.




\section{Results}

The following sections summarize the main results of the analysis of
the various Geant4 physics modelling options; they concern the
longitudinal energy deposition profile.

The lateral energy deposition pattern is also of interest to proton
therapy, and to other experimental applications as well; related major
sources of epistemic uncertainties in the simulation are the models of
multiple scattering and nuclear elastic scattering, and secondary
particle production from nuclear interactions.
The data samples produced for the study of the longitudinal energy
deposition profile are of inadequate size to draw any statistically
significant conclusions concerning the effects of epistemic
uncertainties on the lateral energy distribution patterns; their
quantification at comparable significance level would require simulation samples
at least two orders of magnitude larger for the analysis of
the lateral distribution of energy deposition than for the
longitudinal one.
Such a large scale simulation production in a realistic experimental
use case was beyond the practical reach of the limited computational
resources available to the authors; it should be pursued once adequate
computing means are available.

Unless differently specified in the text, the results were produced
with Geant4 version 9.3.

\subsection{Shaping the  longitudinal energy deposition}
\label{sec_shape}

\begin{figure}
\centerline{\includegraphics[angle=0,width=8.5cm]{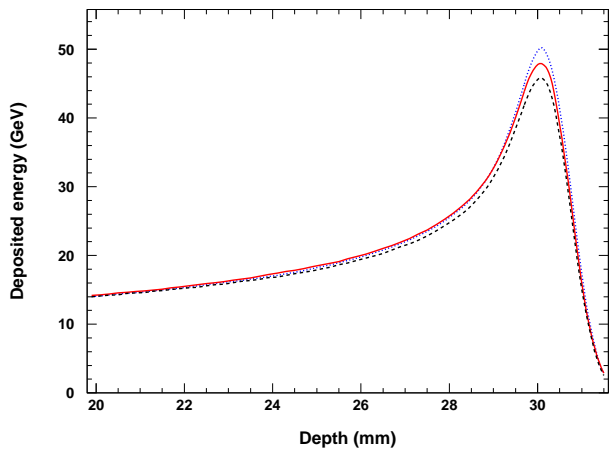}}
\caption{Bragg peak profile resulting from electromagnetic interactions only
(blue dotted curve), electromagnetic interactions and hadronic elastic
scattering (solid red curve), electromagnetic, hadronic elastic and inelastic
scattering (dashed black curve).
The profiles were produced with the Geant4 9.3 physics configuration
listed in Table \ref{tab_refconf} and subsets of it.
The Bragg peak is from one million primary protons,
generated with $\langle E \rangle = 63.95$ MeV mean energy and 
$\sigma = 300$ keV standard deviation incident on water; the plot shows the deposited 
energy in each slice of the longitudinally segmented readout geometry,
integrated over all the generated events. }
\label{fig_emhadro}
\end{figure}

Electromagnetic, hadronic elastic and inelastic interactions contribute to a
different extent to shaping the energy deposition as a function of penetration depth;
hadronic interactions are known \cite{rosenfeld1,rosenfeld2} to be relevant to
the simulation of the proton Bragg peak.
The relative contribution was evaluated by activating partial and full
components of the reference physics configuration in the simulation:
electromagnetic interactions only, electromagnetic interactions and hadronic
elastic scattering, and the full set, including
hadronic inelastic processes, as in Table \ref{tab_refconf}.
Fig. \ref{fig_emhadro} shows the longitudinal energy deposition resulting from
the various contributions: the peak depth and the overall pattern of energy
deposit are dominated by the electromagnetic component, while elastic and
inelastic hadronic interactions contribute to modify its shape.
Other options of Geant4 electromagnetic and hadronic models result in
similar apportioning among the various physics contributions.

The energy deposited in the sensitive volume derives from protons, 
electrons and other particles; the contribution from the latter
amounts to less than 1\%.

\begin{figure}
\centerline{\includegraphics[angle=0,width=8.5cm]{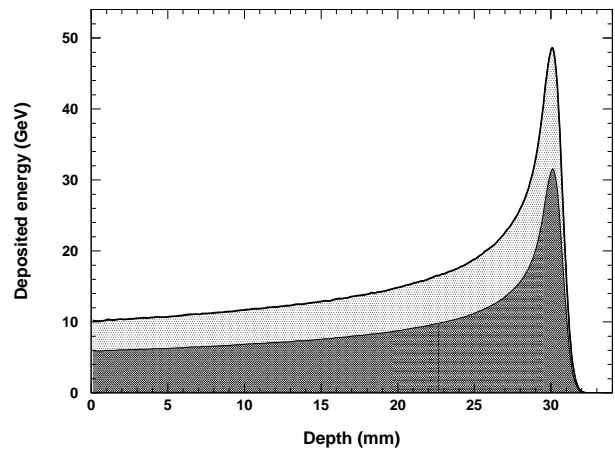}}
\caption{Longitudinal energy deposition due to protons (dark shaded area) and
electrons (light shaded area); the thick line represents the total energy
deposited by all particle species.
The profiles were produced with the Geant4 9.3 physics configuration listed in
Table \ref{tab_refconf}; the electron production threshold was equivalent to 250
eV in water.
The Bragg peak is from one million primary protons
with $\langle E \rangle = 63.95$ MeV and 
$\sigma = 300$ keV incident on water; the plot shows the deposited 
energy in each slice of the longitudinally segmented readout geometry,
integrated over all the generated events.}
\label{fig_electrons}
\end{figure}

The relevance of electrons' contribution to the deposited energy is related to the
electron production threshold set in the simulation application; the results
described in this paper were produced with a threshold
equivalent to 250 eV in the sensitive water volume, that is the lowest energy
recommended for use with Geant4 library-based electron and photon processes, and
corresponds to the setting in the validation study of \cite{sandia_tns}.
The resulting contribution of secondary electrons to the longitudinal energy
deposit profile is illustrated in Fig. \ref{fig_electrons}.

The accuracy of the secondary electron simulation contributes to the overall
accuracy of the energy deposition deriving from primary protons; epistemic
uncertainties in the electron simulation models may affect the results.
For Geant4, the validation of the longitudinal energy deposition
of electrons in an energy range relevant to this study is documented in 
\cite{sandia_tns}.

Low-energy electrons are of great importance in therapeutical particle beams,
since they are very powerful in causing lethal damage to the cells \cite{kempe};
the accuracy of $\delta$-ray simulation is especially important for studies of
the biological effects of proton irradiation.



The passage of primary particles through the beam line affects the acceptance,
i.e. the fraction of primary particles reaching the sensitive volume, and
the characteristics of the particles entering it.

\begin{figure}
\centerline{\includegraphics[angle=0,width=8.5cm]{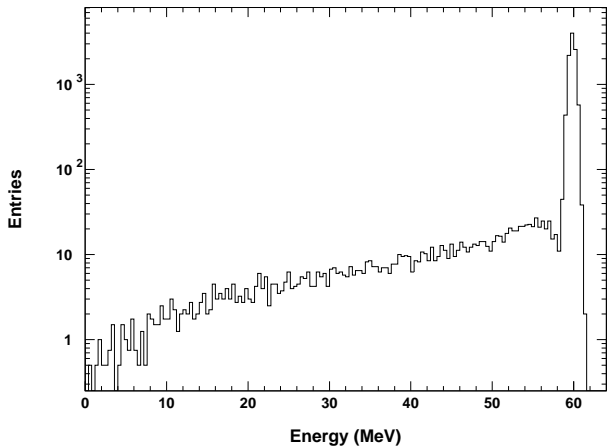}}
\caption{Proton energy spectrum at the entrance of the sensitive water
volume, after traversing the beam line; the primary
beam was modelled according to a gaussian distribution,
with $\langle E \rangle = 63.95$ MeV and $\sigma = 300$ keV.}
\label{fig_psurface}
\end{figure}

\begin{figure}
\centerline{\includegraphics[angle=0,width=8.5cm]{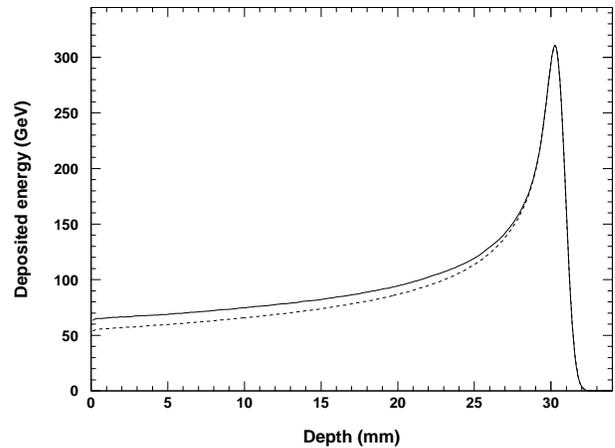}}
\caption{Energy deposition versus penetration depth resulting from the 
whole spectrum of particles
entering the water phantom (solid line) and produced only by the peak portion of
the spectrum of Fig. \ref{fig_psurface} (dashed line), excluding the low energy
tail.
The peak portion of the primary proton spectrum 
at the entrance has 59.823 MeV mean energy and 376 keV 
standard deviation; the figure shows the deposited 
energy in each slice of the longitudinally segmented readout geometry,
integrated over 280000 primary protons entering the sensitive water volume.}
\label{fig_notail}
\end{figure}

The spectrum of protons at the entrance of the sensitive volume, after traversing 
the beam line, is shown in Fig. \ref{fig_psurface}.
The proton interactions in the beam line shift the mean of the energy
distribution to a lower value than the nominal beam energy of 63.95 MeV and
broaden its width, originally of 300 keV: the peak part of the spectrum in
Fig. \ref{fig_psurface} is well fit (with p-value 1) by a gaussian distribution 
with 59.823 MeV mean and 376 keV width (standard deviation).
The energy spectrum of the protons entering the sensitive volume is
characterized by an extended low energy tail, which affects the plateau of the
energy deposition profile in the water volume: low energy protons
stop at lower depth in the water volume, producing a localized large energy
deposit typical of the Bragg peak.
This effect is highlighted in Fig. \ref{fig_notail}, which compares
the energy deposition profile resulting from the proton spectrum shaped
by the beam line to the one produced by the same spectrum, where the
tail was suppressed. 
To suppress the tail, 
primary protons at the entrance of the sensitive volume, whose energy 
differed by more than three standard deviations from the 59.823 MeV 
peak value, were not tracked further.
A data sample consisting of 280000 primary protons entering the 
sensitive volume was used for producing the energy deposition profile in 
Fig. \ref{fig_notail}; this sample is larger than the ones shown in 
other figures to better expose the effects of the low energy proton tail.

In the plateau at lower penetration depth the difference between the two curves
amounts to more than 15\%, while the shape of the peak is hardly affected.
This feature affects parameters used in clinical practice to evaluate
the quality of the irradiation, like the peak to entrance ratio.
This analysis shows that imprecise knowledge of the beam line geometry and
materials can affect the energy deposited in the sensitive volume; for accurate
simulation of the energy deposition in the sensitive water volume, not only
accurate modelling of particle interactions in water is important, but also in
the materials of the beam transport line.

In experimental practice, the features of the particle spectrum should be
taken into account in the choice of the optimal technique for the validation of
simulation models: for instance, Faraday-cup based dosimetry is more sensitive
to the energy distribution of the proton beam than ionization chambers or
calorimeters \cite{icru78_dose}; the presence of a small admixture of low-energy
scattered protons can lead to significant errors in absorbed dose determination
with Faraday cup. \cite{icru78_dose}-\cite{verhey}.

The acceptance values calculated with the various physics model combinations
listed in Table \ref{tab_produced} are all compatible within the statistical
uncertainties.
Further details about the effects of physics models, and their associated
epistemic uncertainties, on the determination of the acceptance in the sensitive
volume are examined in section \ref{sec_mscatt}.

\subsection{Water mean ionisation potential}
\label{sec_ionipot}

Knowledge of the mean excitation energy of a medium is needed to calculate the
energy loss of a charged particle penetrating that medium; various theoretical
calculations and experimental measurements are documented in the literature
concerning this parameter.
The value (75$\pm 3$~eV) recommended in ICRU Report 49 \cite{icru49} is commonly
used in Monte Carlo codes (e.g. Geant4, FLUKA, MCNPX). Nevertheless, values
differing from this reference have recently been proposed: among them, 80.8$\pm
3$~eV in \cite{paul_ionipot}, based on theoretical and experimental
considerations, 61.77 eV in \cite{schulte2004}, 79.7$\pm$0.5~eV in
\cite{bichsel} and 81.6 eV in \cite{dingfelder}; a lower value of 67.2~eV is
assumed in ICRU Report 73 \cite{icru73}, \cite{paul_ionipot}.
An experimental determination of 78.4$\pm$1.0~eV was reported in
\cite{kumazaki}, where a Geant4 simulation encompassing the \emph{ICRU49}
stopping power model was utilized.
The lack of consensus about the value of this parameter corresponds to
an epistemic uncertainty in the simulation.

\begin{figure}
\centerline{\includegraphics[angle=0,width=8.5cm]{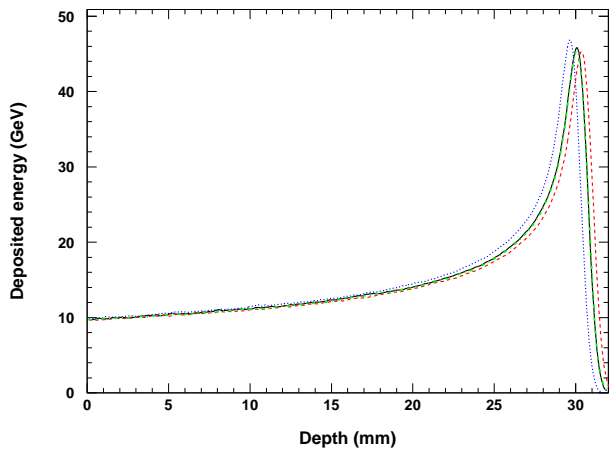}}
\caption{Bragg peak profile resulting from different water ionisation
potentials and proton beam energies: 75~eV \cite{icru49} (solid black curve), 
67.2~eV (dotted blue curve), 80.8~eV  (dashed red curve) with proton
beam energy of 63.95 MeV, and 80.8~eV  (dot-dashed green curve) 
with proton beam energy of 63.65 MeV. 
The Bragg peak is from one million primary protons
with $\langle E \rangle = 63.95$ MeV and 
$\sigma = 300$ keV incident on water; the plot shows the deposited 
energy in each slice of the longitudinally segmented readout geometry,
integrated over all the generated events.}
\label{fig_ionipot}
\end{figure}

The effect of the uncertainty of the water ionization potential was estimated by
performing simulations with values of 75 eV (as in the reference physics
configuration), 67.2 eV and 80.8 eV; apart from this feature, the application
activated the physics configuration summarized in Table \ref{tab_refconf}.
The longitudinal energy deposition profiles corresponding to different values
are shown in Fig. \ref{fig_ionipot}.
A small shift in the depth of the peak is visible; the 67.2 eV and 80.8 eV
settings displace the peak to the adjacent readout geometry slices with respect
to the depth resulting from the water ionization potential set at 75~eV.
As described in section \ref{sec_simu}, the resolution in the Bragg peak
location achievable in the simulation is 200~$\mu m$, which corresponds to the
longitudinal segmentation of the readout geometry.

Similar effects were also observed in simulations with the SHIELD-HIT code
\cite{andreo2009} and with FLUKA \cite{rumeni}; \cite{paganetti2008} reports
approximately 1~mm shift between Geant4-based Bragg peak simulations of 85.6
and 209.2 MeV proton beams, respectively with 75 eV and 70.9 eV water mean
ionization potential (however, without specifying the longitudinal resolution of
the deposited energy collection).

In experimental practice, the ionization potential is usually treated as a
free parameter in the simulation, which is adjusted to improve the match between
experimental and simulated data (e.g. 
\cite{sommerer,goethem,paganetti_parodi,greiner2009}).
It is worth noting that different optimal values of this parameter were
identified in the literature to best match the respective experimental data.

The experimental environment typical of therapeutical beam lines provides
limited insight into this simulation feature, due to the common practice of
empirically determining the optimal beam parameters based on the comparison
between simulated and observed depth dose profiles, as discussed in section
\ref{sec_intro}.
A test was performed to investigate this issue: two simulations were executed
with different, but equally plausible settings - respectively with 63.95 proton
beam energy and 75 eV ionization potential, and with 63.65 MeV proton beam
energy and 80.8 eV ionization potential; the beam energy shift is compatible
with typical uncertainties of the experimental environment under study.
The resulting longitudinal energy deposition profiles, shown in Fig.
\ref{fig_ionipot}, are practically undistinguishable: the peak positions
coincide, and the goodness of fit tests mentioned in section \ref{sec_analysis}
confirm their compatibility at 90\% confidence level.
The comparison with experimental data, whose beam energy is not known a priori
with adequate precision, would not be capable of discriminating the accuracy of
such distributions.

The systematic effects highlighted by this analysis are relevant
only when the simulation is expected to play a predictive role.
In common applications, where the simulation is used only for verification
purpose, the empirical adjudstments of the water mean ionization potential and
of the proton beam parameters mask any potential systematic effects.
The experimental discrimination of the simulation accuracy resulting from
different water ionization potential values would require precise knowledge of
the beam parameters and accurate measurements of the energy deposition as a
function of depth, such that displacements of the Bragg peak smaller than 
200 $\mu m$, associated with the value of this parameter, could be appreciated.

\subsection{Proton stopping powers}
\label{sec_stopping}


Compilations of proton stopping powers are available in \cite{icru49},
\cite{ziegler77}, \cite{ziegler85}, and in the SRIM \cite{srim} code.
Despite the wide body of experimental data, theoretical calculations and
empirical models of proton stopping powers, no consensus has yet been achieved
on definitely established values.
Evaluations of empirical and theoretical stopping power models reported in the
literature \cite{paul_nim227}, \cite{paul_nim247} are limited to a few elements
and compounds; they highlight differences among the various compilations.
According to these analyses, more recent stopping power models do not
necessarily correspond to improved accuracy; some models describe the stopping
powers for some materials well, but appear less accurate for other materials.

Due to this controversial situation, proton stopping powers are a source of
epistemic uncertainty in the simulation results.
A study was performed to evaluate the effects on the Bragg peak profile related
to different stopping power models implemented in Geant4.


The simulations were performed with physics settings as in Table
\ref{tab_refconf}, apart from configuring the low energy hadron ionization
process with various stopping power parameterisation models.
The energy range of application was set according to the recommendations of the
respective references for the \textit{ICRU49}, \textit{Ziegler77} and
\emph{Ziegler85} parameterisation models; lacking specific documentation about
the applicability of the \emph{Ziegler2000} parameterisation, this model was
applied up to 2~MeV, as for \emph{ICRU49}.

\begin{figure}
\centerline{\includegraphics[angle=0,width=8.5cm]{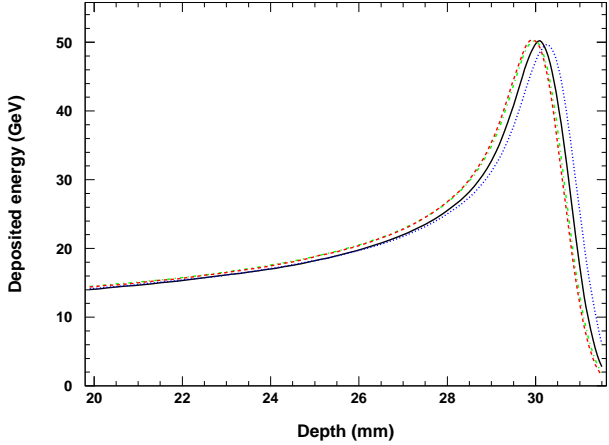}}
\caption{Energy deposition as a function of depth resulting from different proton
stopping power models: \emph{ICRU49} (solid black curve), \emph{Ziegler77}
(dash-dotted green curve) \emph{Ziegler85} (dashed red curve) and
\emph{Ziegler2000} (dotted blue curve); the profiles correspondign to the
\emph{Ziegler77} and \emph{Ziegler85} models are barely distinguishable in the
plot, due to the great similarity of their simulation results.
The Bragg peak is from one million primary protons
with $\langle E \rangle = 63.95$ MeV and 
$\sigma = 300$ keV incident on water; the plot shows the deposited 
energy in each slice of the longitudinally segmented readout geometry,
integrated over all the generated events.}
\label{fig_stopping}
\end{figure}

The results are illustrated in Fig. \ref{fig_stopping}: the various proton
stopping power models produce slightly different energy deposition profiles; the
\emph{Ziegler77} and \emph{Ziegler85} models produce almost identical results.
The peaks associated with alternative proton stopping power models are
located in adjacent longitudinal readout geometry slices with respect
to the peak produced by the reference configuration of Table \ref{tab_refconf}, 
including \emph{ICRU49} stopping powers; as described in section \ref{sec_simu},
the longitudinal readout segmentation is 200~$\mu$m.
Apart from the shift in the peak position, the shapes of the energy deposition
profiles are statistically compatible at 90\% confidence level according to all
the goodness of fit tests mentioned in section \ref{sec_analysis}.

Similarly to the discussion in the previous section concerning the water mean
ionization potential, the epistemic uncertainty related to the stopping power
model used in the simulation turns into systematic effects only when the
simulation is required to play a predictive role; otherwise, as shown in the
previous case, a small adjustment of the proton beam energy, compatible with
typical experimental uncertainties, would shift the energy deposition profiles
deriving from different stopping power models into statistically equivalent
distributions.
These considerations suggest that typical proton therapy experimental
environments would not be sensitive to the differences of stopping
power models, nor would they be adequate to discriminate their accuracy.
All the available stopping power models appear equally suitable to
that simulation environment; in this respect, one can observe that the
use of different Geant4 models has been reported with satisfactory
agreement against experimental data, despite the fact that they determine
different Bragg peak depths: for instance, \emph{Ziegler2000}
in \cite{aso} and \emph{ICRU49} in \cite{pablo,paganetti2004,shipley}.

If predictive capabilities are required from the simulation, higher
precision experimental measurements would be necessary to discriminate
the accuracy of the existing models with the capability of appreciating
shifts in the peak depth smaller than 200 $\mu m$.

This context should be taken into account when considering comparative
evaluations of the accuracy of simulation models:
the procedure of empirically adjusting the parameters in the simulation, based
on a selected physics configuration, to best fit the experimental data is prone
to bias further comparisons of other simulation models with the same data.
The estimate of the relative accuracy of alternative physics models would
require the capability of comparing the simulation outcome to measurements, 
without privileging any of the models to constrain any simulation parameters.

\subsection{Hadronic elastic scattering}
\label{sec_elastic}

Four elastic scattering modeling options available in Geant4 were compared: the
\emph{G4UHadronElasticProcess} with the \emph{G4HadronElastic} model, 
the \emph{G4HadronElasticProcess} process with the \emph{G4LElastic} (LEP)
or \emph{G4ElasticCascadeInterface} (Bertini elastic) models, and the 
CHIPS \emph{G4QElastic} process.

\begin{figure}
\centerline{\includegraphics[angle=0,width=8.5cm]{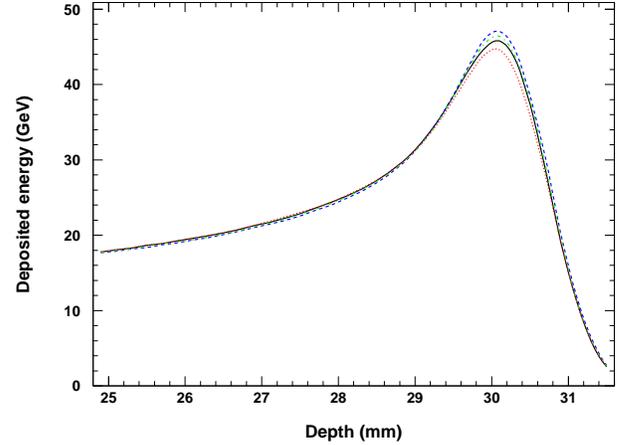}}
\caption{Energy deposition profiles associated with various proton elastic
scattering options: "U-elastic" (solid black curve), Bertini (blue dashed
curve), LEP (green dot-dashed curve) and CHIPS (red dotted curve).
The Bragg peaks are from one million primary protons
with $\langle E \rangle = 63.95$ MeV and 
$\sigma = 300$ keV incident on water; the plot shows the deposited 
energy in each slice of the longitudinally segmented readout geometry,
integrated over all the generated events.}
\label{fig_peak_elastic}
\end{figure}

\begin{figure}
\centerline{\includegraphics[angle=0,width=8.5cm]{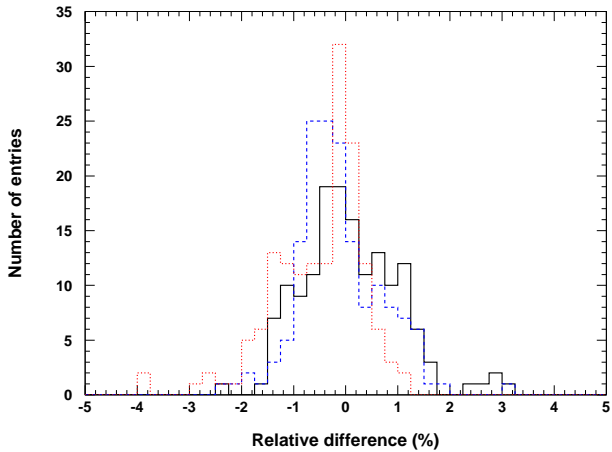}}
\caption{Relative difference of the energy deposited in the longitudinal slices
of the sensitive water volume for various elastic
scattering configurations, with respect to the results of the configuration 
as in Table \ref{tab_refconf} with \emph{G4UHadronElasticProcess} and 
the \emph{G4HadronElastic}
model: Bertini (solid black histogram), LEP (blue dashed histogram) and CHIPS
(red dotted histogram) elastic scattering.
The relative difference is calculated in each slice of the longitudinal
segmentation of the readout geometry associated with the sensitive volume. 
The energy deposition derives from one million primary protons
with $\langle E \rangle = 63.95$ MeV and $\sigma = 300$ keV incident on water.}
\label{fig_diff93_elastic}
\end{figure}

The simulation application activated the physics configuration as in Table
\ref{tab_refconf} for all other features apart from proton elastic
scattering.
The longitudinal energy deposition profiles resulting from the various
simulation configurations are shown in Fig. \ref{fig_peak_elastic}.
The distribution of the relative difference of the energy deposited in
each longitudinal slice of the sensitive volume with respect to the
outcome from the reference configuration of Table \ref{tab_refconf} is
shown in Fig. \ref{fig_diff93_elastic} for the various options; the
differences are mostly comprised within $\pm$2\%.

The compatibility of the energy deposition profiles associated with the 
various elastic scattering options is confirmed
by goodness-of-fit tests, whose results are reported in Table \ref{tab_gofelastic}.
All the tests fail to reject the null hypothesis of compatibility with the
profile deriving from the reference configuration, with 0.1 significance.
A subset of elastic scattering modeling options, limited to the ``U-elastic'',
``Bertini'' and ``LEP'' ones, was compared in the context of Geant4 8.1p02 and 9.1
versions as well.
All the considered elastic scattering options were compatible with 0.1
significance within a given Geant4 version; the results concerning the
comparison of the left branch of the energy deposition profiles are reported in
Table \ref{tab_gofelastic}.

If the differences between the outcome of two simulation configurations were
due only to statistical fluctuations, one would expect them to be
distributed, as a function of depth, in a rather large number of short 
sequences (runs) of consecutive positive and negative values;
the Wald-Wolfowitz test \cite{wald} was applied to evaluate this hypothesis.
The resulting p-value is smaller than 0.001 for all the test
cases; therefore one can infer some systematic effects associated with
the choice of the elastic scattering model in the simulation.
It is worth remarking that the conclusion drawn from the Wald-Wolfowitz test
does not contradict the result of the goodness-of-fit tests: the two types
of tests, respectively evaluating the differences between two distributions in
terms of sign and of distance, are complementary.
A feature of the energy deposition profiles hinting at
systematic differences is visible in the vicinity of the Bragg peak in Fig.
\ref{fig_peak_elastic}, where alternative elastic scattering options appear
associated with sequences of energy deposition consistently larger or smaller
than those deriving from the configuration of Table \ref{tab_refconf} encompassing
the "U-elastic" elastic scattering model.


For the use case under study, the small differences exhibited by the
various simulation models look compatible with the experimental
resolution typical of the application domain (of the order of 2-2.5\% 
\cite{icru78_dose}); therefore, the peculiarities
of the models do not affect the outcome of the simulation significantly.
Based on these results, one can conclude that at the present stage all
the Geant4 elastic scattering options are equivalent for 
the use case considered in this study.
Validation against experimental data concerning the energy range
and target materials pertinent to this use case would strengthen the 
predictive reliability of the simulation.

\begin{table}
\begin{center}
\caption{P-value of goodness-of-fit tests comparing hadronic 
elastic scattering options}
\label{tab_gofelastic}
\begin{tabular}{|l|l|l|c|c|c|}
\hline
Version			& Range	& Model		& Kolmogorov 	& Anderson	& Cramer 	\\
			&    	&		& Smirnov	& Darling 	& von Mises 	\\
\hline
\multirow{9}{*}{9.3}	&	& Bertini	& 1		& 1		& 1		\\
			& Whole	& LEP		& 1		& 1		& 1		\\
			&	& CHIPS		& 0.997		& 0.997		& 0.999		\\
\cline{2-6}
			& Left	& Bertini	& 1		& 1		& 1		\\
			&branch	& LEP		& 1		& 1		& 1		\\
			&	& CHIPS		& 0.996		& 0.982		& 0.999		\\
\cline{2-6}
			&Right	& Bertini	& 1		& 0.986		& 0.972		\\
			&branch	& LEP		& 1		& 0.986		& 0.972		\\
			&	& CHIPS		& 1		& 0.986		& 0.972		\\
\hline
9.1			& Left	& Bertini	& 1		& 1		& 1		\\
			&branch	& LEP		& 0.996		& 0.989		& 1		\\
\hline
8.1			& Left	& Bertini	& 1		& 0.999		& 0.997		\\
			&branch	& LEP		& 1		& 1		& 1		\\
\hline
\end{tabular}
\end{center}
\end{table}


\subsection{Hadronic inelastic cross sections}

The proton total inelastic cross sections are an important parameter in the
simulation of therapeutical proton beams, since they determine the amount of
proton loss from the primary beam.

Two configurations of cross sections were evaluated in this study: those
implemented in \emph{G4HadronInelasticDataSet}, originating from GHEISHA,
and those implemented in \emph{G4ProtonInelasticCrossSection} and
\emph{G4NeutronInelasticCrossSection}, respectively for protons and
neutrons, covering the energy range above 6.8 MeV.
Apart from this feature, all the other physics options in the simulation were
set as in Table \ref{tab_refconf}.

Both cross sections derive from parameterisations of experimental data;
it is not clear whether the comparisons available in the literature 
concern the calibration of the parameterisation with experimental data, 
or represent the cross section model validation.

No significant dependence on the cross section options is observed
regarding the proton acceptance in the sensitive water volume, which
is affected by the interactions to which primary protons are subject
in the materials of the beam line.

The two sets of cross sections determine some difference in the occurrence
of the proton inelastic scattering process associated with them in the
sensitive water volume.
Confidence intervals for this quantity were calculated, using Student's $t$
distribution, based on the simulation sample activating the cross sections as in
\cite{axen}.
The 99\% confidence interval for the mean value of hadronic
inelastic scattering occurrences lies between 1688 and 1849, when one million
primary protons are generated, while the number of occurrences with the
GHEISHA-like cross section data set is 1654;
this value is significantly different from the mean number of inelastic
scattering occurrences determined by the cross sections of \cite{axen}.

\begin{figure}
\centerline{\includegraphics[angle=0,width=8.5cm]{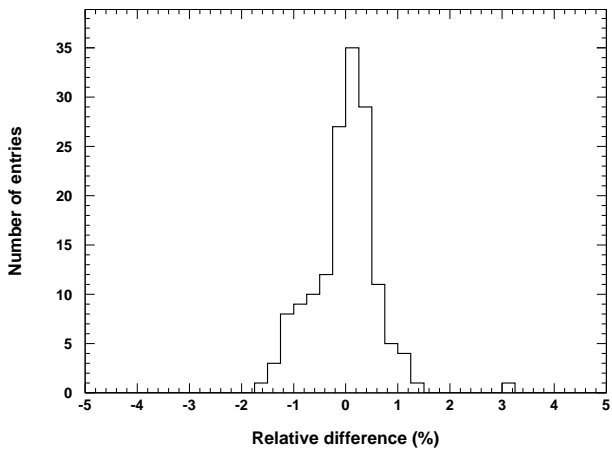}}
\caption{Percent relative difference of the energy deposition profiles
resulting from the activation of 
\emph{G4HadronInelasticDataSet} hadronic inelastic cross sections or
\emph{G4ProtonInelasticCrossSection} and \emph{G4NeutronInelasticCrossSection}; 
all the other simulation options are identical in the two cases, and set as in
Table \ref{tab_refconf}.
The energy deposition derives from one million primary protons
with $\langle E \rangle = 63.95$ MeV and $\sigma = 300$ keV incident on water.
The relative difference is calculated in each slice of the longitudinal
segmentation of the readout geometry associated with the sensitive volume. }
\label{fig_cross}
\end{figure}

Nevertheless, the effect of this difference on the longitudinal energy 
deposition appears negligible.
The distribution of the relative differences of the energy deposition profiles
associated with the two options is shown in Fig. \ref{fig_cross}; it is
consistent with typical experimental uncertainties in hadron therapy practice.
The longitudinal energy deposition profiles resulting from the two cross section
options, with other physics settings as in Table \ref{tab_refconf},
are compatible at 90\% confidence level according to all the
aforementioned goodness-of-fit-tests.

Therefore one can conclude that the characteristics of the two hadronic
cross section data sets do not affect the simulation of the proton depth dose
profiles in the use case considered in this study.


\subsection{Hadronic inelastic scattering models}

Several alternative hadronic inelastic scattering modeling options were
evaluated: the Precompound model, the Bertini and  Li\`ege cascade models, 
the LEP parameterised model and the CHIPS model.
In addition, a few configuration options of the nuclear deexcitation phase,
accessible through the interface of the \emph{G4ExcitationHandler} class
instantiated by the Precompound model, were evaluated together with the
Precompound model: the generalized evaporation (GEM) model replacing the default 
evaporation one, and the activation of the Fermi break-up.
The Precompound model was evaluated both as a standalone model and as
invoked by the Binary cascade model.
 
Proton and neutron interactions were handled consistently in each simulation
configuration by activating the same hadronic inelastic model option for both
particles; all the other physics features were set as in Table
\ref{tab_refconf}.

The longitudinal energy deposition profiles produced by the various 
hadronic inelastic models in Geant4 9.3 appear visually undistinguishable;
therefore no related figure is shown in this paper.

\begin{figure}
\centerline{\includegraphics[angle=0,width=8.5cm]{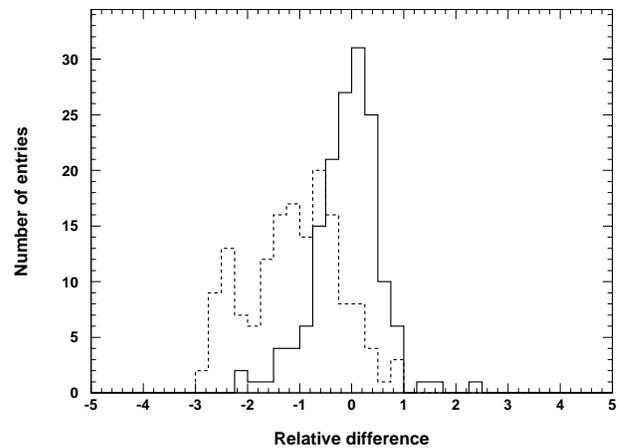}}
\caption{Relative difference of longitudinal energy deposition profiles
associated with the GEM evaporation model, with respect to the ``reference
physics configuration'': simulation based on Geant4 9.3 (solid line) and 9.1
(dashed line).
The observed systematic effect is related to the correction of a software
feature in Geant4 9.3.
The energy deposition derives from one million primary protons
with $\langle E \rangle = 63.95$ MeV and $\sigma = 300$ keV incident on water.
The relative difference is calculated in each slice of the longitudinal
segmentation of the readout geometry associated with the sensitive volume.  }
\label{fig_gem}
\end{figure}

The energy deposition profile produced with the GEM evaporation model closely
resembles the one deriving from the default evaporation model instantiated in
connection with the Precompound model, as seen in Fig. \ref{fig_gem};
this observation is confirmed by the results of the goodness-of-fit tests in
Table \ref{tab_gofinelastic} with 0.1 significance.
Differences related to the use of the two models were visible with
previous Geant4 versions, as shown in Fig. \ref{fig_gem};
the GEM implementation was subject to improvements in Geant4 9.3
\cite{relnotes93}.

\begin{figure}
\centerline{\includegraphics[angle=0,width=8.5cm]{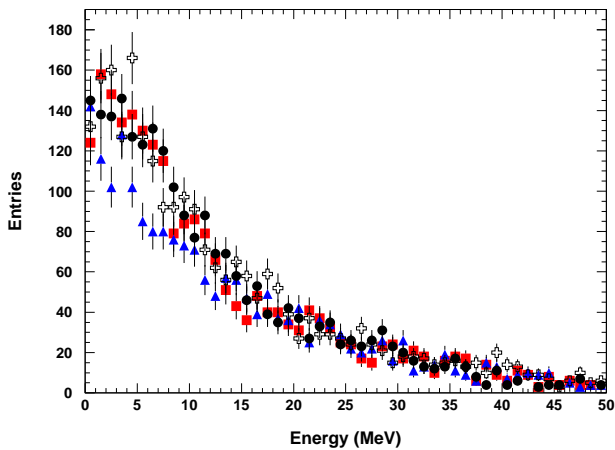}}
\caption{Energy spectrum of secondary protons produced with different
configurations of the Precompound model: default configuration as in Table
\ref{tab_refconf} (black circles), configuration with GEM evaporation (red
squares), configuration activating Fermi break up (blue triangles) and
configuration activating the Binary Cascade model (white crosses), which in turn
invokes the Precompound model to handle the preequilibrium phase. 
The secondary spectra derive from one million primary protons
with $\langle E \rangle = 63.95$ MeV and 
$\sigma = 300$ keV incident on water.}
\label{fig_protons_gem93}
\end{figure}

The distributions of the secondary particles produced in association with the
two evaporation models look consistent, compatible with statistical fluctuations;
the secondary proton distributions are shown in Fig. \ref{fig_protons_gem93}
as an example.
The lack of visible effects does not necessarily mean that these two models are
characterized by identical features; rather, it shows that the use case under
study is not sensitive to their possible difference.

From this analysis one can conclude that the evaporation model options are
equivalent for the simulation of the longitudinal energy deposition; as documented
in section \ref{sec_cpu}, the GEM model is computationally faster than the
default one in the application under study.

The evaporation process of nuclear deexcitation is based
on the hypothesis that the excitation energy is high and approximately
equally distributed among the nucleons: this assumption 
is justified for heavy nuclei,
but it is not applicable to the water target considered in this use case.
It is generally accepted that the Fermi break-up represents a more appropriate
theoretical description of the nuclear deexcitation process for light nuclei:
in MCNPX and FLUKA the Fermi break-up is applied to nuclei with atomic mass 
up to 17, whereas evaporation is applied to heavier nuclei;
in Geant4 it is not invoked by default in the deexcitation of
light nuclei, although the public interface of \emph{G4ExcitationHandler} allows
users to modify the default settings.

\begin{figure}
\centerline{\includegraphics[angle=0,width=8.5cm]{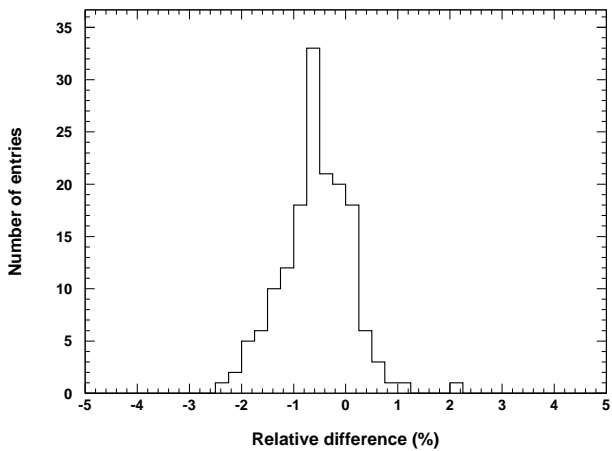}}
\caption{Relative difference of the longitudinal energy deposition profile
deriving from the activation of the Fermi break-up, with respect to the 
profile deriving from the default 
configuration of the Precompound model; all the other settings
are as in Table \ref{tab_refconf}.
The energy deposition derives from one million primary protons
with $\langle E \rangle = 63.95$ MeV and $\sigma = 300$ keV incident on water.
The relative difference is calculated in each slice of the longitudinal
segmentation of the readout geometry associated with the sensitive volume. }
\label{fig_differmi}
\end{figure}

The effects of the Fermi break-up were evaluated by activating it in the
simulation for nuclei with atomic number smaller than 10; they are 
visible in the spectrum of the produced secondaries, with
respect to those produced by nuclear deexcitation proceeding through evaporation.
From a theoretical perspective, the application of an evaporation model 
to the deexcitation of light nuclei
is expected to overestimate the production of secondary protons in the 
lower energy range; this effect is indeed observed in Fig. \ref{fig_protons_gem93}.
The activation of the Fermi break up affects the longitudinal energy deposition,
with respect to that resulting from the default nuclear deexcitation settings:
their relative difference, shown in Fig. \ref{fig_differmi}, exhibits an asymmetric 
distribution shifted towards negative values. 
This effect hints at a systematic contribution of the Fermi break-up to decrease
the longitudinal energy deposition; nevertheless, the observed differences are
consistent with typical uncertainties in experimental proton therapy
practice.


\begin{figure}
\centerline{\includegraphics[angle=0,width=8.5cm]{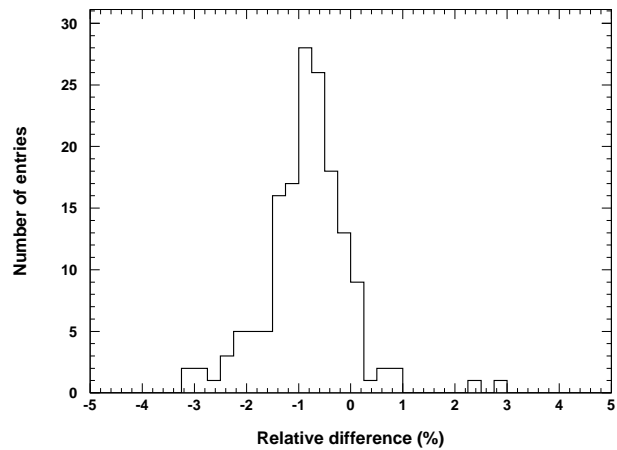}}
\caption{Relative difference of the longitudinal energy deposition profile deriving
the activation of the Precompound model either as a standalone model, or as
invoked by the Binary cascade model.
The energy deposition derives from one million primary protons
with $\langle E \rangle = 63.95$ MeV and $\sigma = 300$ keV incident on water.
The relative difference is calculated in
each slice of the longitudinal segmentation of the readout geometry associated
with the sensitive water volume. }
\label{fig_diffbinary}
\end{figure}

A similar asymmetry in the longitudinal energy deposition difference is observed
when the preequilibrium phase is associated with intranuclear transport; this
effect is shown in Fig. \ref{fig_diffbinary}, which concerns two simulations
involving the Precompound model, respectively as an independent model and
invoked through the Binary Cascade model.
The transition between the cascade process of intranuclear transport and the
preequilibrium is determined by empirical considerations \cite{binary}, which
are specific to each software implementation: for instance, in Geant4 Binary
cascade model cascading continues as long as there are particles above a 70 MeV
kinetic energy threshold \cite{binary} (along with other conditions required by
the algorithm), while a smooth transition around 50 MeV is implemented in FLUKA
\cite{fluka1}.
The empirical features of the algorithm correspond to lack of knowledge from
physical principles to determine the transition between the two r\'egimes; the
analysis shows that this epistemic uncertainty can be a source of systematic 
effects, such as the bias of the distribution in Fig. \ref{fig_diffbinary}.
This effect, which is of a magnitude comparable to typical experimental
uncertainties in hadron therapy measurements, could be significant in use cases
where precise predictive power is expected from the Monte Carlo simulation.

No such asymmetries, with respect to simulating the preequilibrium
phase only (as in the Precompound model), are observed with two other
configurations involving intranuclear cascade models - the Bertini and
Li\`ege ones.
It is worth remarking that the Li\`ege model does not involve a preequilibrium
phase at all, while the Bertini cascade model does. 
The apparent absence of consistent trends related to the adopted physical
approach (modeling intranuclear cascade, preequilibrium and their interplay)
suggests that the observed behavior of the code may be influenced by other
implementation details on top of the basic physics modeling approach; this
consideration adds further complexity to the effort of identifying the sources
of epistemic uncertainty in the simulation, which is a necessary step towards
their reduction or their control.


\begin{figure}
\centerline{\includegraphics[angle=0,width=8.5cm]{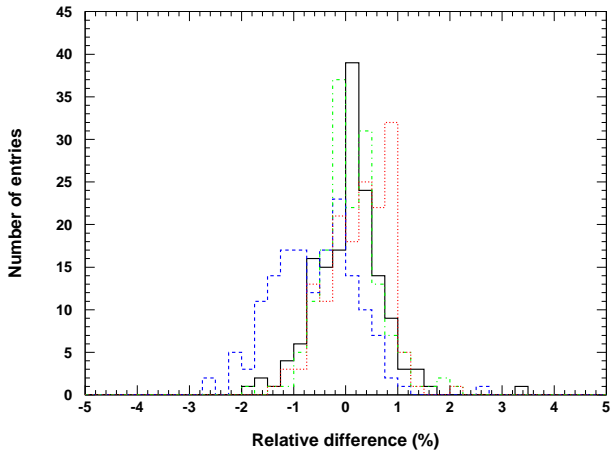}}
\caption{Relative difference of the energy deposited in the longitudinal slices
of the sensitive water volume associated with various hadronic inelastic
options, with respect to the configuration of Table \ref{tab_refconf} 
encompassing the
Precompound model: Bertini (solid black histogram), LEP (blue dashed histogram),
Li\`ege (green dot-dashed histogram) and CHIPS (red dotted histogram).
The energy deposition derives from one million primary protons
with $\langle E \rangle = 63.95$ MeV and $\sigma = 300$ keV incident on water.}

\label{fig_diff93_inelastic}
\end{figure}

The relative differences of the energy deposition profiles concerning other
hadronic inelastic models with respect to the outcome deriving from 
the Precompound one are illustrated in Fig. \ref{fig_diff93_inelastic}.

All the final state models of hadronic inelastic scattering produce
statistically compatible results at 90\% confidence level in the considered use case, 
as shown in Table \ref{tab_gofinelastic}.

The Wald-Wolfowitz test concerning the difference with respect to the reference
physics configuration results in a p-value smaller than 0.001 for all the
considered modeling options, with the exception of the comparison involving the
Li\`ege cascade model, for which the p-value is 0.360.
These results suggest the presence of some systematic effects due to the choice 
of the hadronic inelastic models in the simulation; some asymmetries and
bias with respect to zero are indeed visible in the distributions in Fig.
\ref{fig_diff93_inelastic}, namely the one concerning the LEP inelastic model,
apart from the previously discussed effects in Fig. \ref{fig_differmi} and
\ref{fig_diffbinary}.

Like the results discussed in section \ref{sec_elastic}, this result suggests that
the selection of the hadronic inelastic model activated in the simulation can be
source of systematic effects.
Nevertheless, the differences concerning the longitudinal energy deposition
patterns appear compatible with typical experimental uncertainties in proton
therapy dosimetry; therefore the systematic effects identified by the
Wald-Wolfowitz test would be negligible in that software application context.
They could become relevant in use cases where higher accuracy is demanded.


\begin{table*}
\begin{center}
\caption{P-value of goodness-of-fit tests comparing hadronic 
inelastic scattering options}
\label{tab_gofinelastic}
\begin{tabular}{|c|l|l|c|c|c|}
\hline
{\bf Geant4}	&{\bf Test}	&{\bf Hadronic}	&{\bf Kolmogorov} 	&{\bf Anderson}	&{\bf Cramer} 	\\
{\bf version}	&{\bf range}    &{\bf model}	&{\bf Smirnov}	&{\bf Darling} 	&{\bf von Mises} \\
\hline
\multirow{9}{*}{\bf 9.3}	&	& Bertini	& 1	& 1		& 1		\\
			& 	& LEP		& 0.954		& 0.988		& 0.984 	\\
			& Whole	& Li\`ege	& 1		& 1		& 1		\\
			&	& CHIPS		& 1		& 1		& 1		\\
			&	& GEM		& 1		& 1		& 1		\\
			&	&Fermi break-up & 1		& 1		& 1		\\
			&	& Binary	& 0.954		& 0.938		& 0.973		\\
\cline{2-6}
			&	& Bertini	& 1		& 1		& 1		\\
			& Left	& LEP		& 0.945		& 0.961		& 0.979	 	\\
		& branch	& Li\`ege	& 1		& 1		& 1		\\
			&	& CHIPS		& 1		& 1		& 1		\\
			&	& GEM		& 1		& 1		& 1		\\
			&	&Fermi break-up & 1		& 1		& 1		\\
			&	& Binary	& 0.945		& 0.858		& 0.962		\\
\cline{2-6}
			&	& Bertini	& 1		& 0.986		& 0.972	\\
			& Right	& LEP		& 1		& 0.986		& 0.972	\\
		& branch	& Li\`ege	& 1		& 0.986		& 0.972	\\
			&	& CHIPS		& 1		& 0.986		& 0.972	\\
			&	& GEM		& 1		& 0.986		& 0.972	\\
			&	&Fermi break-up & 1		& 0.986		& 0.972		\\
			&	& Binary	& 1		& 0.986		& 0.972		\\
\hline
{\bf 9.1}		& Left	& Bertini	& 0.981		& 0.901		& 0.980	\\
		& branch	& LEP		& 0.945		& 0.949		& 0.937	\\
\hline
{\bf 8.1}		& Left	& Bertini	& 1		& 1		& 1		\\
			&branch	& LEP		& 0.996		& 0.814		& 0.847	\\
\hline
\end{tabular}
\end{center}
\end{table*}

The implementation of the Geant4 Precompound model was subject to improvements
\cite{quesada} in Geant4 9.2.
Nevertheless, these modifications do not appear to have affected the features of
the longitudinal energy deposition pattern significantly, since the goodness-of-fit
tests in Table \ref{tab_gofinelastic} show that the energy deposition
profiles associated with this model were compatible with those deriving from
other hadronic inelastic models in previous Geant4 versions, as well as in the
9.3 one.
This remark is relevant to previous applications of the Precompound model 
to the use case under study, which are archived in the literature.
	
\begin{figure}
\centerline{\includegraphics[angle=0,width=8.5cm]{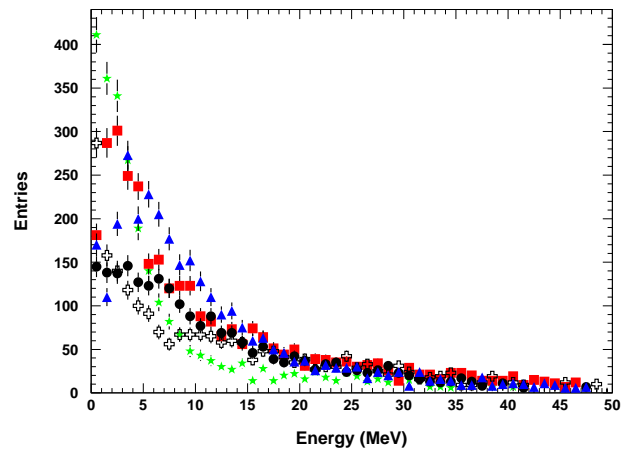}}
\caption{Secondary proton spectrum resulting from various hadronic inelastic
models: Precompound (black circles), Bertini (red squares), LEP (white crosses),
Li\`ege (blue triangles) and CHIPS (green stars).
The secondary particle spectrum derives from one million primary protons
with$\langle E \rangle = 63.95$ MeV and 
$\sigma = 300$ keV incident on water.}
\label{fig_second_p}
\end{figure}

\begin{figure}
\centerline{\includegraphics[angle=0,width=8.5cm]{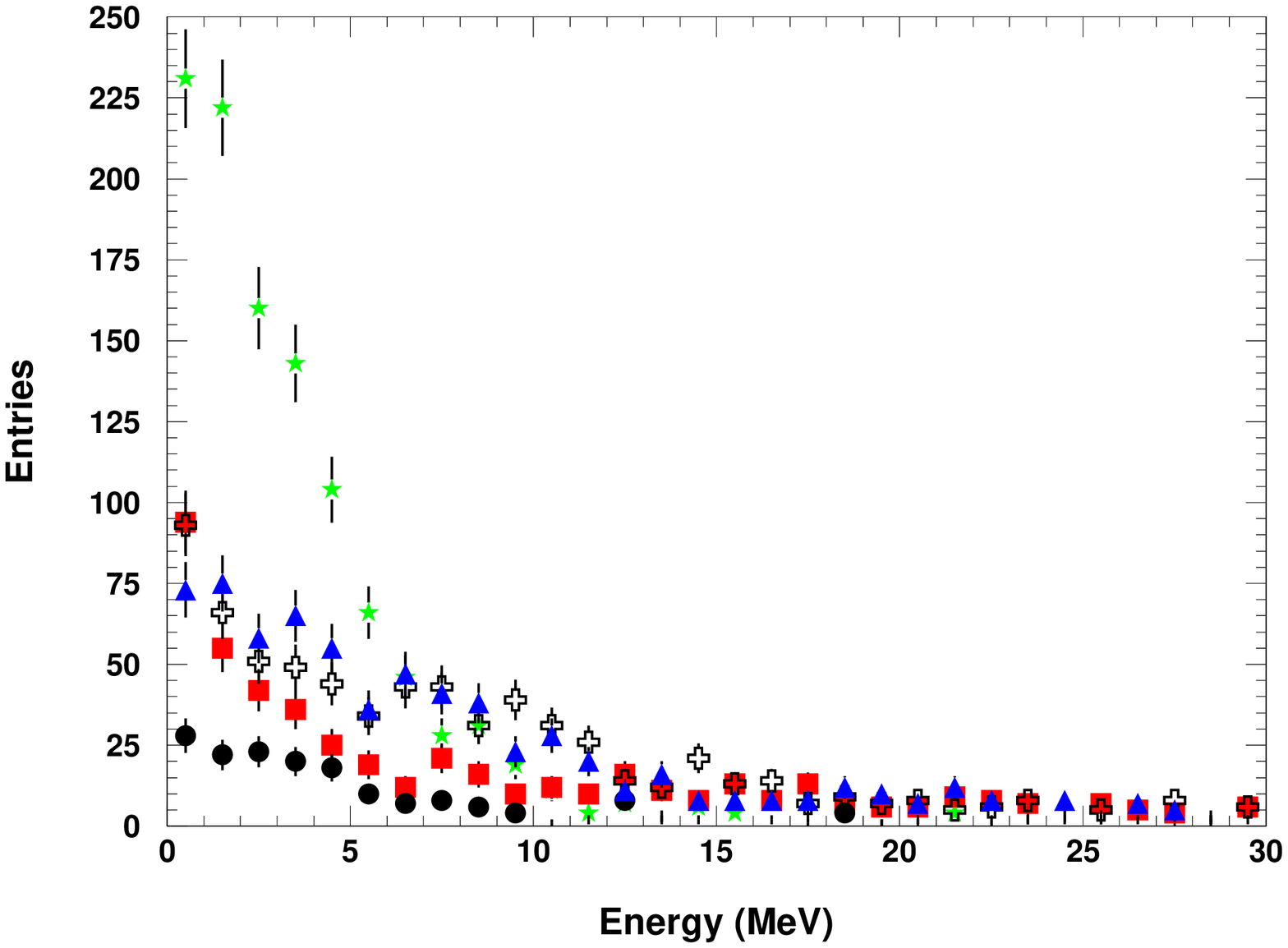}}
\caption{Secondary neutron spectrum resulting from various hadronic inelastic
models: Precompound (black circles), Bertini (red squares), LEP (white crosses),
Li\`ege (blue triangles) and CHIPS (green stars).
The secondary particle spectrum derives from one million primary protons
with $\langle E \rangle = 63.95$ MeV and 
$\sigma = 300$ keV incident on water.}
\label{fig_second_n}
\end{figure}

\begin{figure}
\centerline{\includegraphics[angle=0,width=8.5cm]{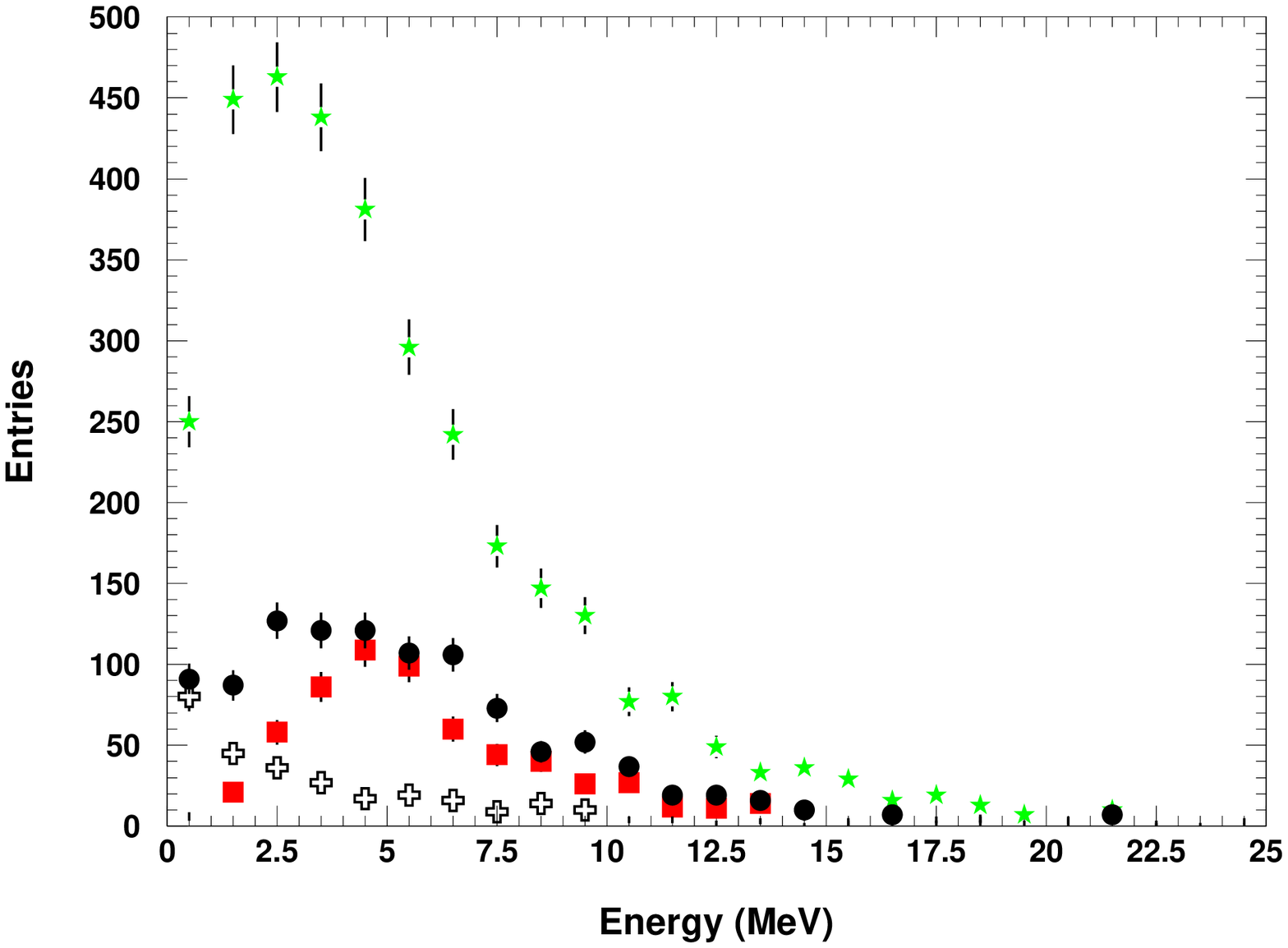}}
\caption{Secondary $\alpha$ particle spectrum resulting from various hadronic inelastic
models: Precompound (black circles), Bertini (red squares), LEP (white crosses)
and CHIPS (green stars); no $\alpha$ particles appear
to be produced by the Geant4 implementation of the Li\`ege model.
The secondary particle spectrum derives from one million primary protons
with $\langle E \rangle = 63.95$ MeV and 
$\sigma = 300$ keV incident on water.}
\label{fig_second_alpha}
\end{figure}

Nevertheless, despite their similarity at determining the longitudinal
energy deposition profile, some hadronic inelastic models exhibit very
different characteristics regarding the secondary particles they generate:
the secondary proton, neutron and $\alpha$ particle spectra are shown in
Fig. \ref{fig_second_p} to \ref{fig_second_alpha}.
Radiotherapy applications can be affected by secondary particles
within the target volume and outside it, both laterally and beyond the distal
edge of the Bragg peak \cite{icru78_beam}; concerns for the risks due to 
secondary particles in proton therapy are discussed in the literature \cite{paganetti_n}.
The analysis documented in the previous paragraphs shows that the 
different secondary particle production patterns do not produce significant
effects on the longitudinal energy deposition profile; the quantification of 
possible effects on the lateral energy deposition pattern would require
substantially larger data samples, which were not achievable with the 
limited computational resources available to the authors in the course 
of the project documented in this publication, and is outside the scope 
of this paper.

The identification of actual systematic effects related to hadronic inelastic
models, and their quantitative estimate, 
would require experimental measurements with adequate accuracy to
discriminate not only the features of the energy deposition distribution,
but also the characteristics of the secondary particles they produce.


\subsection{Multiple Coulomb scattering}
\label{sec_mscatt}


The configuration of proton 
multiple scattering simulation in a Geant4 application
involves the selection of the multiple scattering process and models to be
activated, and setting some parameters used by the multiple 
scattering algorithm.
Default options are provided in Geant4 kernel for the model and parameters
associated to multiple scattering processes; they are summarized in Table
\ref{tab_mspara} for a set of recent Geant4 versions.

\begin{table*}
\begin{center}
\caption{Default settings of relevant multiple scattering processes}
\label{tab_mspara}
\begin{tabular}{|c|l|c|c|c|c|c|l|}
\hline
Geant4	&Process		&Range	&Step	&Lateral	&skin   &Geometry	&Model \\
version	&			&Factor	&Limit	&Displacement	&	&Factor		& \\
\hline
8.1	&G4MultipleScattering	&0.02	&1	&		&	&		&Urban \\
9.1	&G4MultipleScattering	&0.02	&1	&1		&0	&2.5		&Urban \\
9.2p0.3	&G4MultipleScattering	&0.02	&1	&1		&3	&2.5		&Urban \\
9.3	&G4MultipleScattering	&0.04	&1	&1		&3	&2.5		&Urban92 \\
9.3	&G4hMultipleScattering	&0.2	&0	&1		&3	&2.5		&Urban90 \\
\hline
\end{tabular}
\end{center}
\end{table*}

The analysis evaluated whether recent evolutions of the code between Geant4
8.1 and 9.3 versions, which involve different model and parameter settings,
could be the source of systematic effects in Geant4-based
simulations for proton therapy applications, originating from epistemic
uncertainties in the simulation model.
Two issues were addressed: the effects related to different multiple scattering
processes, \textit{G4MultipleScattering} and \textit{G4hMultipleScattering}, and
those due to changes in the \textit{G4MultipleScattering} process since the 8.1
release.
It should be remarked that in the following analysis the behaviour associated
with multiple scattering may result not only directly from the
implementation of the two above mentioned classes, but also from 
behavior inherited from their base classes or acquired through aggregation of,
or dependence on, other classes, as determined by the software design of 
Geant4 multiple scattering domain.
The results are reported for Geant4 versions 8.1p02, 9.1, 9.2p03,
and 9.3.

Two multiple scattering processes,
\textit{G4hMultipleScattering} and \textit{G4MultipleScattering},
are applicable to protons in Geant4 9.3.
The former was first introduced in Geant4 8.2 to provide faster
simulation of hadron transport; the latter complies with an earlier
class interface and is planned to be withdrawn from later releases. 
In Geant4 9.3 the common base class \textit{G4VMultipleScattering}
accounts for public member functions formerly specific to
\textit{G4MultipleScattering}.
\textit{G4hMultipleScattering} can be configured to acquire equivalent
behavior to \textit{G4MultipleScattering} by applying the same
settings (model and parameters) as in \textit{G4MultipleScattering}
listed in Table \ref{tab_mspara}.
For convenience, the configuration of \textit{G4hMultipleScattering}
equivalent to \textit{G4MultipleScattering} is still indicated as
\textit{G4MultipleScattering} in the following.


Only the default settings listed in Table \ref{tab_mspara}
were tested; 
the large effects related to epistemic uncertainties observed in this
limited interval analysis, which are documented in the following,
suggest that this complex problem domain would benefit from a larger-scale 
dedicated study beyond the scope of this paper.


To acquire sound evidence of effects associated with
multiple scattering settings, the comparisons were performed over five physics
configurations:
the set of processes and models (apart from multiple scattering) as in 
Table \ref{tab_refconf}, and variants of it consisting of ``LEP'' and ``Bertini''
inelastic scattering together with ``U-elastic'' elastic scattering, 
and ``LEP'' and ``Bertini'' elastic scattering together with the Precompound
hadronic inelastic model.
Common effects observed in such an extended set of test cases could be 
reasonably associated with the multiple scattering domain, excluding 
their possible origin from intrinsic features of a single physics configuration.

\begin{figure}
\centerline{\includegraphics[angle=0,width=8.5cm]{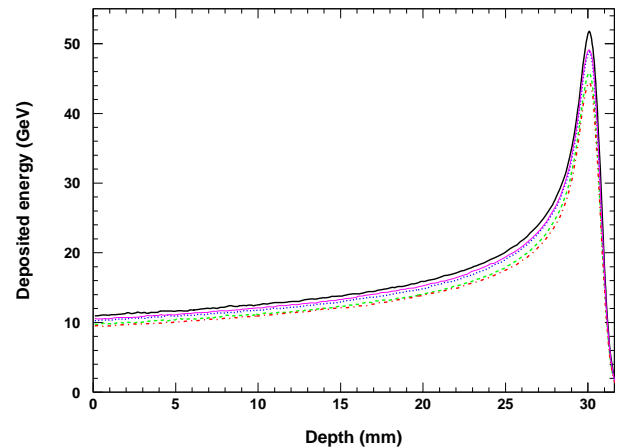}}
\caption{Bragg peak profile resulting from different multiple scattering
processes and Geant4 versions: \textit{G4hMultipleScattering} (black, thick solid line) 
in Geant4 9.3,
\textit{G4MultipleScattering} in Geant4 9.3 (dashed, green line),
Geant4 9.2p03 (pink, thin solid line),
Geant4 9.1 (dotted, blue line)
and Geant4 8.1p02 (dash-dotted, red line) 
The same physics configuration was activated in all the simulations,
apart from the multiple scattering setting.
The Bragg peak is from one million primary protons
with $\langle E \rangle = 63.95$ MeV and 
$\sigma = 300$ keV incident on water; the plot shows the deposited 
energy in each slice of the longitudinally segmented readout geometry,
integrated over all the generated events.}
\label{fig_hmscatt}
\end{figure}

\begin{figure}
\centerline{\includegraphics[angle=0,width=8.5cm]{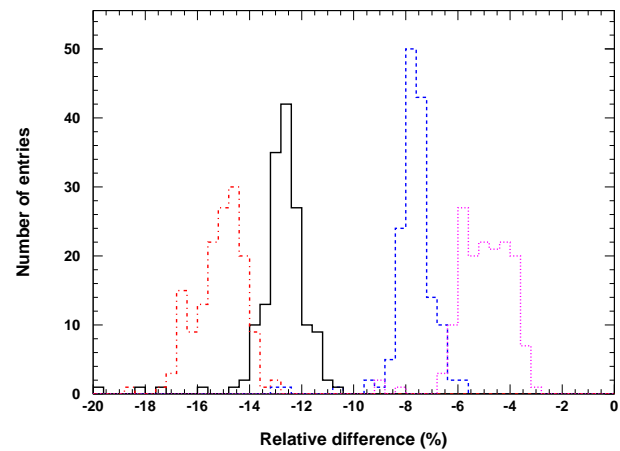}}
\caption{Relative difference of the energy deposited in the longitudinal slices
of the sensitive water volume associated with different multiple scattering 
processes and Geant4 versions; the difference is calculated in 
each longitudinal slice of the readout geometry with respect 
to a reference configuration with \textit{G4hMultipleScattering} in Geant4 9.3
for identical configurations with \textit{G4MultipleScattering} 
in Geant4 9.3 (solid black histogram), 9.1 (dashed blue histogram), 
9.2p03 (dotted pink histogram) and 8.1p02 (dash-dotted red histogram) versions.
The reference configuration is as in Table \ref{tab_refconf}, except for the 
hadronic inelastic scattering option (Bertini instead of Precompound); this
replacement is due to the greater stability of the Bertini code across the 
various Geant4 versions, nevertheless all other physics configurations produce 
similar results.
The energy deposition derives from one million primary protons
with $\langle E \rangle = 63.95$ MeV and $\sigma = 300$ keV incident on water.}
\label{fig_diffverhms}
\end{figure}

The resulting longitudinal energy deposition distributions associated with 
proton multiple scattering options are shown in Fig. \ref{fig_hmscatt}.
The two multiple scattering processes produce visibly different longitudinal
energy deposition profiles; the extent of the differences can be quantitatively
appraised in Fig. \ref{fig_diffverhms}, which shows the variation of the
longitudinal energy deposition profiles simulated with
\textit{G4hMultipleScattering} in Geant4 9.3 with respect to equivalent
simulations performed with \textit{G4MultipleScattering} in Geant4 9.3, 9.1 and
8.1p02.
The plot shows results produced with the Bertini hadronic inelastic option;
similar results are obtained with the other physics configurations subject to
comparison.
The energy deposition profile of Geant4 9.2p03 is not shown in 
Fig. \ref{fig_diffverhms} to avoid clogging the plot with many curves;
it lies in between the profiles produced by Geant4 9.3 with 
 \textit{G4hMultipleScattering} and Geant4 9.1.

The results of goodness-of-fit tests comparing longitudinal energy deposition distributions
associated with either proton multiple scattering process are summarized in
Table \ref{tab_gofhms}.
The longitudinal energy deposition distributions associated with
\textit{G4hMultipleScattering} are incompatible at 99.9\% confidence level with
those produced by Geant4 versions 8.1p02 and 9.3 with \textit{G4MultipleScattering}, 
and, apart from one
test case involving the Kolmogorov-Smirnov test, at 99\% confidence level
with those produced by Geant4 9.1.
Regarding the comparison with the profiles generated with Geant4 9.2p03,
the Anderson-Darling test rejects the hypothesis of compatibility with
the profiles produced with \textit{G4hMultipleScattering}at 95\% confidence 
level, the Kolmogorov-Smirnov does not reject it, while the Cramer-von Mises
rejects it in two physics configurations and does not reject it in the
other three configuration. 
The Anderson-Darling and Cramer-von Mises tests are considered more powerful
than the Kolmogorov-Smirnov test;
the Anderson-Darling test is especially sensitive to fat tails \cite{stephens}.
\textit{G4MultipleScattering} was responsible of the multiple scattering process
in these earlier Geant4 versions.

It is worth mentioning that \cite{paganetti2008} shows
comparisons of experimental and simulated proton energy deposition profiles
normalized to the number of protons in the beam; the reported simulations were
performed with Geant4 8.1p01 version.
The small plots in logarithmic scale and the limitation of the comparisons to
qualitative appraisal prevent the reader from understanding whether the different
behaviour of \textit{G4hMultipleScattering}, and of the
\textit{G4MultipleScattering} class in later Geant4 versions, with respect to
the multiple scattering implementation of Geant4 8.1, would affect the
compatibility with the experimental data of \cite{paganetti2008}.

\begin{figure}
\centerline{\includegraphics[angle=0,width=8.5cm]{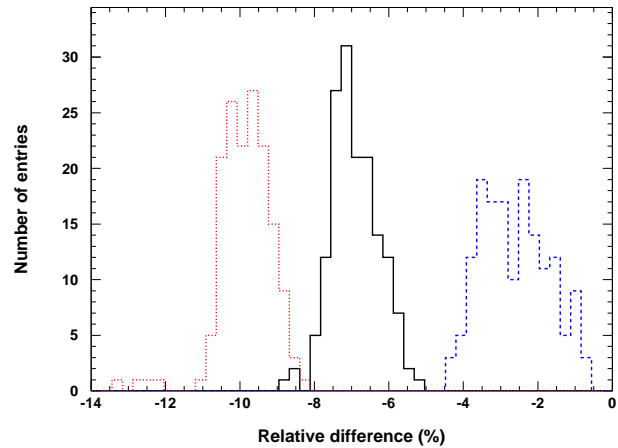}}
\caption{Relative difference of the energy deposited in the longitudinal slices
of the sensitive water volume associated with the
\textit{G4MultipleScattering} in Geant4 9.2p03, with respect to identical
simulation settings in Geant4 9.3 (solid black histogram),
Geant4 9.1 (dashed blue histogram), and Geant4 8.1p02 (dotted red histogram).
The reference configuration is as in Table \ref{tab_refconf}, except for the 
hadronic inelastic scattering option (Bertini instead of Precompound); this
replacement is due to the greater stability of the Bertini code across the 
various Geant4 versions, nevertheless all other physics configurations produce 
similar results.
The energy deposition derives from one million primary protons
with $\langle E \rangle = 63.95$ MeV and $\sigma = 300$ keV incident on water.}
\label{fig_diffver}
\end{figure}
\begin{figure}
\centerline{\includegraphics[angle=0,width=9cm]{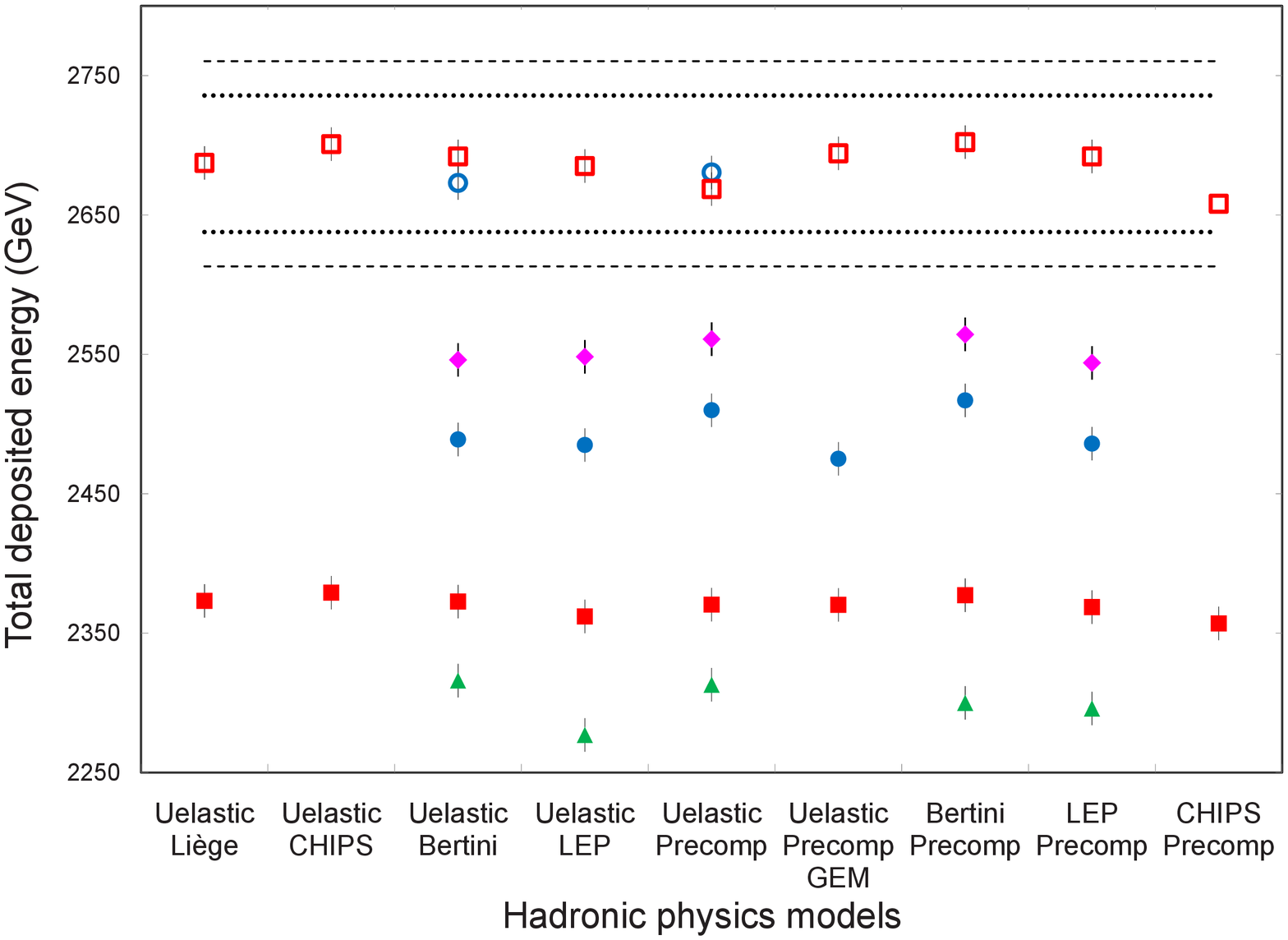}}
\caption{Total energy deposition in the sensitive volume associated with various
Geant4 versions and physics configurations: Geant4 9.3 (red squares), 
Geant4 9.2p03 (pink diamonds), Geant4 9.1
(blue circles) and Geant4 8.1p02 (green triangles); the filled symbols
correspond to simulations activating the \textit{G4MultipleScattering} multiple
scattering process, while the empty ones correspond to the activation of
\textit{G4hMultipleScattering}. The upper and lower lines of the horizontal axis
identify respectively the hadronic elastic and inelastic scattering model in
each simulation configuration; the other physics options, apart from
the multiple scattering under test, were as in Table \ref{tab_refconf}.
The dashed and dotted lines represent respectively 
the 99.9\% and 99\% confidence intervals for the mean value 
of the total deposited energy 
over various Geant4 9.3 physics configurations associated with
\textit{G4hMultipleScattering}. 
The total energy deposition derives from one million primary protons
with $\langle E \rangle = 63.95$ MeV and $\sigma = 300$ keV incident on water.}
\label{fig_etot}
\end{figure}

Visible differences are also observed in Fig. \ref{fig_diffver} concerning
the energy deposition profiles associated with 
\textit{G4MultipleScattering} settings over the various versions.
The total energy deposition shown in Fig. \ref{fig_etot} exhibits evident
differences associated with the various settings.


The Geant4 Low Energy Electromagnetic package, used in all the simulations, was
subject to configuration and Change Management discipline \cite{swprocess}
based on the Unified Software Development Process \cite{up} framework until
Geant4 release 9.1; the adopted software process ensured that the software of
this package relevant to the use case under study did not undergo modifications
between the 8.1p02 and 9.1 production versions, which could alter its physical
behaviour.
The same implementations of the low energy electromagnetic processes were used
in the simulations based on Geant4 9.2 and 9.3; therefore, it is plausible that
variations observed in the simulation productions based on different Geant4
versions are associated with evolutions in other Geant4 domains.
The extent of the differences observed when comparing two Geant4 versions
appears to be approximately the same over all the hadronic physics configurations
activated in the simulation; since the occurrence of coherent modifications to
all the Geant4 hadronic elastic and inelastic scattering implementations is not
likely, this observation suggests that coherent differences would derive 
from modifications either to Geant4 transport kernel or to the multiple
scattering domain, which are common to all the simulations.  
Major changes to Geant4 kernel have not been documented over the 
considered versions; the multiple scattering domain, which was subject to
evolution, appears the most likely source of the observed discrepancies.
Their origin is probably from epistemic uncertainties in the simulation models,
whose validation in the energy range relevant to this use case is scarcely 
documented in literature.


The differences concerning multiple scattering settings
in the various Geant4 versions are significant.
The 99.9\% and 99\% confidence intervals for the mean value of the total energy deposition
deriving from \textit{G4hMultipleScattering} in Geant4 9.3 are shown in Fig.
\ref{fig_etot}; the values deriving from Geant4 versions 8.1p02, 9.1 and 9.2p03 fall
outside the  99.9\% confidence interval.

The results of goodness-of-fit tests are reported in  Table \ref{tab_gofhms}.
In most test cases the longitudinal energy deposition distributions produced 
with Geant4 9.3 are incompatible with those produced with Geant4 9.1 at 95\%
confidence level; in a few cases the test statistic results in p-values close 
to the critical region for 0.05 significance. 
The null hypothesis of equivalence of the distributions subject to test is not
rejected, with the same significance, in the comparisons involving Geant4 9.3
and 8.1p02 versions.
This quantitative result is consistent with the qualitative appraisal of Fig.
\ref{fig_hmscatt}, where the energy deposition profile deriving from Geant4 9.3
appears closer to the one produced with the earlier 8.1p02 version.
The energy deposition profiles produced with Geant4 9.2p03 are incompatible
with those produced with Geant4 9.3 (with \textit{G4MultipleScattering}) 
and Geant4 8.1p02 with 0.05 significance, while the goodness-of-fit tests fail to 
reject the hypothesis of compatibility with the profiles produced with
Geant4 9.1 with 0.05 significance.
The longitudinal energy deposition distributions produced with Geant4 9.1 and 
8.1p02 are incompatible with 0.05 significance.

The simulations with the two multiple scattering processes and with different
Geant4 versions produce a significantly different total 
energy deposition in the sensitive volume.
The results are shown in Fig. \ref{fig_etot}; the dashed and dotted lines in the plot
represent respectively the 99.9\%  and 99\%
confidence intervals for the average energy deposition with
\textit{G4hMultipleScattering} over all Geant4 9.3 physics configurations.

The absolute value of the energy deposition is relevant to applications where
knowledge of the actual dose released to a target is critical, like oncological
treatment planning, radiation protection or radiation damage estimate.
The observed differences would be significant in use cases where the simulation
has a predictive role: differences greater than 10\% in the dose
released to a patient, like the effects observed with the various multiple
scattering implementations released in Geant4, would be important in clinical
applications.
These use cases would not be limited to the bio-medical application domain; for
instance, the use of Monte Carlo simulation to study the damage to electronic
components exposed to radiation would require precise estimate of the released
dose.

\begin{figure}
\centerline{\includegraphics[angle=0,width=9cm]{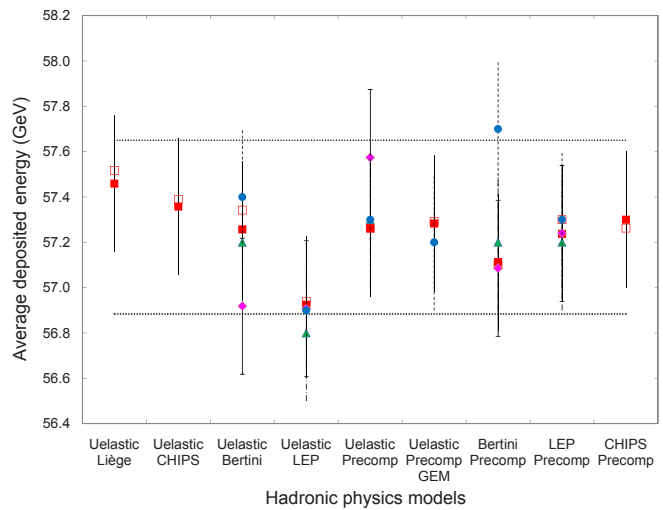}}
\caption{Average energy deposition per proton entering the sensitive volume
calculated with various Geant4 versions and physics configurations: Geant4 9.3
(red squares), Geant4 9.2p03 (pink diamonds), 
Geant4 9.1 (blue circles) and Geant4 8.1p02 (green triangles);
the filled symbols correspond to simulations activating the
\textit{G4MultipleScattering} multiple scattering process, while the empty ones
correspond to the activation of \textit{G4hMultipleScattering}. The upper and
lower lines of the horizontal axis identify respectively the hadronic elastic
and inelastic scattering model in each simulation configuration; the other
physics options, apart from the multiple scattering under test, were as in Table
\ref{tab_refconf}.
The dotted lines represent the 95\% confidence interval for the mean value 
of the various Geant4 9.3 physics configurations associated with
\textit{G4hMultipleScattering}.
The average energy deposition derives from one million primary protons
with $\langle E \rangle = 63.95$ MeV and $\sigma = 300$ keV incident on water.}
\label{fig_eavg}
\end{figure}

\begin{figure}
\centerline{\includegraphics[angle=0,width=9cm]{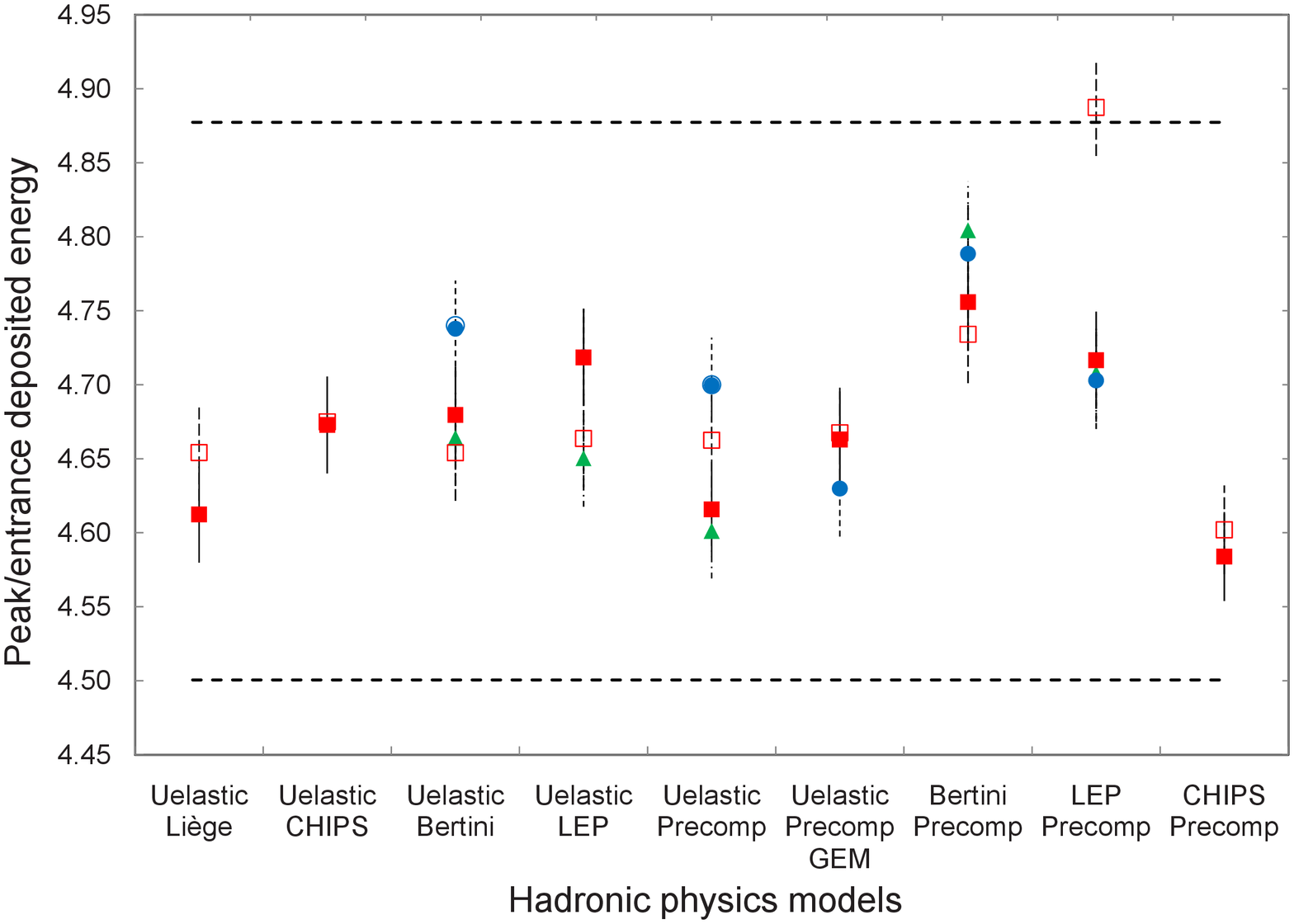}}
\caption{Ratio of the energy deposition at the Bragg peak location and at the
entrance of the sensitive volume, deriving from various Geant4 versions and
physics configurations: Geant4 9.3 (red squares), 
Geant4 9.2p03 (pink diamonds), Geant4 9.1 (blue circles) and
Geant4 8.1p02 (green triangles); the filled symbols correspond to simulations
activating the \textit{G4MultipleScattering} multiple scattering process, while
the empty ones correspond to the activation of \textit{G4hMultipleScattering}.
The upper and lower lines of the horizontal axis identify respectively the
hadronic elastic and inelastic scattering model in each simulation
configuration; the other physics options, apart from the multiple scattering
under test, were as in Table \ref{tab_refconf}.
The dashed lines represent the 95\%
confidence interval for the mean value of the various Geant4 9.3 physics
configurations associated with \textit{G4hMultipleScattering}.
The energy deposition derives from one million primary protons
with $\langle E \rangle = 63.95$ MeV and $\sigma = 300$ keV incident on water.}
\label{fig_peakentrance}
\end{figure}

The average energy deposition per proton in the sensitive volume, and the ratio
between the energy deposited at the peak location and at the entrance of the
sensitive volume are approximately the same for all the physics configurations
and Geant4 versions, as illustrated in Fig. \ref{fig_eavg} and
\ref{fig_peakentrance}. 
The 95\% confidence interval for the mean value deriving
from Geant4 9.3 with \textit{G4hMultipleScattering}
is shown in the figures to appreciate quantitatively the spread
of the results.
These observations suggest that the detailed features of the energy deposition in the
water volume are insensitive to the physics options selected in the simulation,
including multiple scattering, and to the evolutions of Geant4 software.

\begin{figure}
\centerline{\includegraphics[angle=0,width=9cm]{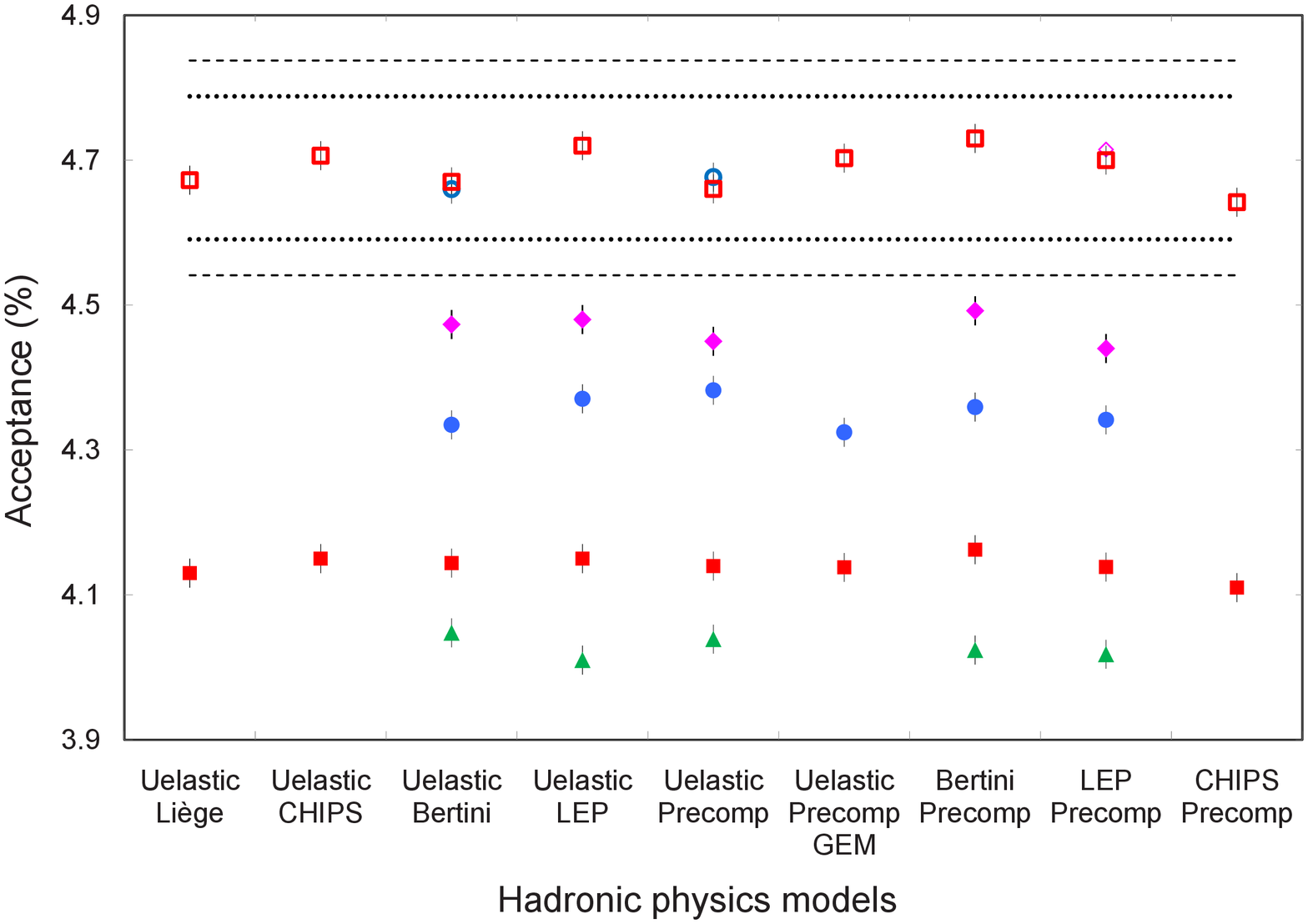}}
\caption{Percentage of primary protons (acceptance) reaching the sensitive
volume deriving from various Geant4 versions and physics configurations: Geant4
9.3 (red squares), Geant4 9.2p03 (pink diamonds), 
Geant4 9.1 (blue circles) and Geant4 8.1p02 (green
triangles); the filled symbols correspond to simulations activating the
\textit{G4MultipleScattering} multiple scattering process, while the empty ones
correspond to the activation of \textit{G4hMultipleScattering}. The upper and
lower lines of the horizontal axis identify respectively the hadronic elastic
and inelastic scattering model in each simulation configuration; the other
physics options, apart from the multiple scattering under test, were as in Table
\ref{tab_refconf}.
The dashed and dotted lines represent respectively the 99.9\% and 99\%
confidence intervals for the mean value
of the various Geant4 9.3 physics configurations of associated with
\textit{G4hMultipleScattering}.
The simulation involves from one million primary protons
with $\langle E \rangle = 63.95$ MeV and $\sigma = 300$ keV incident on water.}
\label{fig_acceptance}
\end{figure}

The acceptance, i.e. the fraction of protons reaching the sensitive volume, out
of all the primary generated ones, is plotted in Fig. \ref{fig_acceptance} for
different physics configurations and Geant4 versions.
Various sources can affect it: inelastic nuclear reactions,
which remove protons from the beam prior to reaching the sensitive volume,
and nuclear elastic and multiple Coulomb scattering, which modify the protons'
direction along with their passage through matter.

The acceptance appears roughly constant in Fig. \ref{fig_acceptance} for the
various hadronic models, within the set of simulations associated with a given
multiple scattering option and Geant4 version; therefore, the features of these
models can be excluded as a source of significant differences.
Complementary tests, whose results are not reported in Fig. \ref{fig_acceptance},
show that the acceptance is not significantly sensitive 
to alternative stopping power models either.
The multiple scattering algorithm appears the most probable source of the
observed differences.


Fig. \ref{fig_etot} and \ref{fig_acceptance} suggest a correlation between the 
total energy deposited in the sensitive volume and the acceptance. 
This effect was evaluated by means of Pearson's correlation coefficient 
\cite{pearson_correlation}; the
null hypothesis consists of assuming no correlation between these quantities.
The correlation coefficient, calculated over all
physics configurations and Geant4 versions examined in this study, is 0.965; the
null hypothesis is rejected with 0.0001 significance.
On the other hand, no correlation of the total energy deposition is observed with
the average energy deposited per proton, nor with the peak over entrance ratio:
the corresponding correlation coefficients are 0.120 and 0.151;
these values lead to not rejecting the null hypothesis with 0.1 significance.

These results hint that the observed discrepancies in the longitudinal energy
deposit distributions are related to effects due to multiple scattering in the
beam line, rather than to physics modeling effects in the water volume.
Geant4 multiple scattering implementation encompasses various empirical
parameters \cite{elles}, whose settings are characterized by epistemic uncertainties;
presumably, the observed effects are associated with different 
angular distribution (including backscattering) and lateral displacement of the
scattered particle implemented in the various Geant4 multiple scattering options
and versions, and the variations of empirical parameters governing the algorithm.

This finding stresses the importance of accurately modeling the beam line
geometry and material composition for accurate calculation of the energy
deposited in the sensitive volume. It also highlights the importance of
correctly simulating particle interactions not only in the sensitive parts of the
experimental set-up, but also in its passive components, since the latter appear to be
responsible for significant systematic effects on the energy deposited in the
sensitive volume.

In hadron therapy practice, proton depth dose profiles are usually normalized
to a reference value (at the peak or at the entrance of the sensitive volume),
or to the integral of the dose, as discussed in \cite{bragg_norma}. 
The same goodness-of-fit tests reported in Table
\ref{tab_gofhms} were performed after normalizing the energy deposition profiles to
the total energy collected in the sensitive volume: they failed to reject the
null hypothesis of compatibility with 0.1 significance, which was rejected in
the comparison of the original (non normalized) distributions.
This analysis demonstrates that normalized distributions are insensitive 
to the large differences exhibited by the various models on an absolute
scale.
Therefore, comparisons like the one in Fig. 7 of \cite{standard_chep07},
concerning an experimental Bragg peak and one simulated with Geant4 9.0, 
both normalized to 1, are of limited usefulness to clarify the issues that emerged in 
the previous analysis.

Further tests were performed, activating the specific
\textit{G4eMultipleScattering} process for electron multiple scattering,
released in Geant4 9.3; no
effects were observed on the longitudinal energy deposition.

The authors found only limited documentation 
in the literature concerning the experimental validation of
proton multiple scattering implementations in recent Geant4 versions; the
comparisons with experimental data reported in \cite{ms_chep2009}
concern muons and electrons and are not pertinent to the use case
object of this investigation, which concerns protons.
Therefore, this process is a source of epistemic uncertainty in the simulation;
the analysis described in this paper shows that this uncertainty could determine
large systematic effects in critical use cases.
Further experimental measurements would be useful for the validation of 
Geant4 multiple scattering in use cases similar to the one considered in this paper;
in particular, experimental data suitable to clarify the interplay between 
energy deposition measurements and features of the multiple scattering algorithm
would be beneficial.  
Ideally, the experiment should be able to measure effects related to
backscattering and lateral displacement, which could be responsible for the
discrepancies in the proton acceptance observed with the various algorithm
implementations examined in this paper.


\begin{table*}
\begin{center}
\caption{P-value of goodness-of-fit tests comparing longitudinal energy deposition
profiles deriving from various Geant4 versions and physics configurations}
\label{tab_gofhms}
\begin{tabular}{|l|l|c|c|c|}
\hline
{\bf Compared} 			&{\bf Physics}		&{\bf Kolmogorov} 	&{\bf Anderson}	&{\bf Cramer} 	\\
{\bf software versions}		&{\bf configuration}	&{\bf Smirnov}	&{\bf Darling} 	&{\bf von Mises} \\	
\hline
{\bf 9.3} (G4hMultipleScattering) 	& Uelastic Bertini	& $<0.001$	& $<0.001$	& $<0.001$	 \\
{\bf 9.3} (G4MultipleScattering) 	& Uelastic LEP		& $<0.001$	& $<0.001$	& $<0.001$	 \\
					& Uelastic Precompound	& $<0.001$	& $<0.001$	& $<0.001$	 \\
					& Bertini Precompound	& $<0.001$	& $<0.001$	& $<0.001$	 \\
					& LEP Precompound	& $<0.001$	& $<0.001$	& $<0.001$	 \\
\hline
{\bf 9.3} (G4hMultipleScattering) 	& Uelastic Bertini	& 0.219		& 0.046		& 0.110	 \\
{\bf 9.2} (G4MultipleScattering)	& Uelastic LEP		& 0.074		& 0.013		& 0.047	 \\
					& Uelastic Precompound	& 0.170		& 0.037		& 0.111	 \\
					& Bertini Precompound	& 0.054		& 0.012		& 0.044	 \\
					& LEP Precompound	& 0.098		& 0.019		& 0.057	 \\
\hline
{\bf 9.3} (G4hMultipleScattering) 	& Uelastic Bertini	& 0.009	& 0.001		& 0.004	 \\
{\bf 9.1} (G4MultipleScattering)	& Uelastic LEP		& 0.002	& $<0.001$	& 0.002	 \\
					& Uelastic Precompound	& 0.014		& 0.002		& 0.008	 \\
					& Bertini Precompound	& 0.004		& $<0.001$	& 0.002	 \\
					& LEP Precompound	& 0.006		& 0.001		& 0.006	 \\
\hline
{\bf 9.3} (G4hMultipleScattering) 	& Uelastic Bertini	& $<0.001$	& $<0.001$	& $<0.001$	 \\
{\bf 8.1} (G4MultipleScattering)	& Uelastic LEP		& $<0.001$	& $<0.001$	& $<0.001$	 \\
					& Uelastic Precompound	& $<0.001$	& $<0.001$	& $<0.001$	 \\
					& Bertini Precompound	& $<0.001$	& $<0.001$	& $<0.001$ \\
					& LEP Precompound	& $<0.001$	& $<0.001$	& $<0.001$	 \\
\hline
{\bf 9.3} (G4MultipleScattering)	& Uelastic Bertini	& 0.006		& $<0.001$	& 0.004 \\
{\bf 9.2} (G4MultipleScattering)	& Uelastic LEP		& 0.001		& $<0.001$	& 0.002	 \\
					& Uelastic Precompound	& 0.006		& $<0.001$	& 0.003	 \\
					& Bertini Precompound	& $<0.001$	& $<0.001$	& 0.001	 \\
					& LEP Precompound	& 0.009		& $<0.001$	& 0.005	 \\
\hline
{\bf 9.3} (G4MultipleScattering)	& Uelastic Bertini	& 0.277	& 0.051		& 0.113 	\\
{\bf 9.1} (G4MultipleScattering)	& Uelastic LEP		& 0.039	& 0.011		& 0.043	 \\
					& Uelastic Precompound	& 0.054		& 0.012		& 0.040	 \\
					& Bertini Precompound	& 0.039		& 0.007		& 0.028	 \\
					& LEP Precompound	& 0.130		& 0.030		& 0.071	 \\
\hline
{\bf 9.3} (G4MultipleScattering)	& Uelastic Bertini	& 0.803	& 0.505		& 0.722	 \\
{\bf 8.1} (G4MultipleScattering)	& Uelastic LEP		& 0.277	& 0.119		& 0.270	 \\
					& Uelastic Precompound	& 0.515		& 0.232		& 0.475	 \\
					& Bertini Precompound	& 0.219		& 0.072		& 0.179	 \\
					& LEP Precompound	& 0.426		& 0.150		& 0.297	 \\
\hline
{\bf 9.2} (G4MultipleScattering)	& Uelastic Bertini	& 0.426	& 0.135		& 0.286	 \\
{\bf 9.1} (G4MultipleScattering)	& Uelastic LEP		& 0.709	& 0.281		& 0.395	 \\
					& Uelastic Precompound	& 0.884 & 0.418		& 0.548	 \\
					& Bertini Precompound	& 0.709	& 0.324		& 0.478	 \\
					& LEP Precompound	& 0.426	& 0.269		& 0.516	 \\
\hline
{\bf 9.2} (G4hMultipleScattering) 	& Uelastic Bertini	& $<0.001$	& $<0.001$	& $<0.001$	 \\
{\bf 8.1} (G4MultipleScattering)	& Uelastic LEP		& $<0.001$	& $<0.001$	& $<0.001$	 \\
					& Uelastic Precompound	& $<0.001$	& $<0.001$	& $<0.001$	 \\
					& Bertini Precompound	& $<0.001$	& $<0.001$	& $<0.001$ \\
					& LEP Precompound	& $<0.001$	& $<0.001$	& $<0.001$	 \\
\hline
{\bf 9.1} (G4MultipleScattering)	& Uelastic Bertini	& 0.020		& 0.004		& 0.016	 \\
{\bf 8.1} (G4MultipleScattering)	& Uelastic LEP		& 0.001		& $<0.001$	& 0.001	 \\
					& Uelastic Precompound	& 0.003		& $<0.001$	& 0.003	 \\
					& Bertini Precompound	& $<0.001$	& $<0.001$	& $<0.001$	 \\
					& LEP Precompound	& 0.006		& $<0.001$	& 0.003	 \\
\hline
\end{tabular}
\end{center}
\end{table*}

\section{Computational performance}
\label{sec_cpu}

The extensive survey of physics models and parameters relevant to the problem
domain documented in the previous sections provides guidance for Geant4-based
simulations concerning similar use cases.
The computational performance of the available physics options is a relevant
parameter in simulation applications, especially considering that some 
previous analyses demonstrate the equivalence of some of them regarding the
physical features they produce.
Therefore the analysis is complemented here by some information on the associated
computational performance in the use case described in this paper, which
can be useful to experimentalists in their Geant4-based applications.

The results reported in Table \ref{tab_cpu}, related to Geant4 9.3 version,
show the average simulation time per primary generated event in each physics
configuration; they derive from the productions for the analysis described in
the previous sections.
The content of Table \ref{tab_cpu} should not be considered as measurements of
Geant4 computational performance in absolute terms: the application code
contained analysis features, such as filling a large number of histograms, which
added an additional burden to the execution with respect to the time strictly
needed for particle transport; moreover no effort was invested in the
optimization of the application code.
However, since all the simulations reported in that table were run on identical
hardware and platforms, the measured execution times are interesting for
relative comparisons of the computational performance of the various physics
configurations in the use case object of this study.

The results reported in Table \ref{tab_cpu} involve the \textit{ICRU49} proton
stopping power model; simulations involving the \textit{Ziegler77},
\textit{Ziegler85} and \textit{Ziegler2000} models are slightly slower.
Simulations involving \textit{G4hMultipleScattering} require approximately 5\%
more CPU time than those involving \textit{G4MultipleScattering}; however, the
larger acceptance associated with this multiple scattering model requires longer
computations to track a greater number of particles in the sensitive volume.

It is worth remarking that accounting for nuclear interactions in the simulation
application described in this paper increases the computational time consumption
by approximately 57\%, with respect to considering electromagnetic interactions
only.

\begin{table}
\begin{center}
\caption{Averge CPU time per primary generated event in various physics configurations}
\label{tab_cpu}
\begin{tabular}{|l|l|c|}
\hline 
\textbf{Hadronic elastic} 		& \textbf{Hadronic inelastic} 			& \textbf{CPU time (ms)}  	\\
\hline
Bertini-elastic		& Precompound				& 254.0 	$\pm$ 0.3 \\
LEP				& Precompound				& 255.1 	$\pm$ 0.3 \\
CHIPS-elastic		& Precompound				& 293.3 	$\pm$ 0.3 \\
U-elastic			& Precompound				& 254.3 	$\pm$ 0.3 \\
U-elastic			& Precompound-GEM			& 251.1 	$\pm$ 0.3 \\
U-elastic  		         & Precompound-Fermi break-up 	& 255.8	$\pm$ 0.3 \\
U-elastic		         & Binary cascade   			& 261.3	$\pm$ 0.3 \\
U-elastic			& Bertini cascade				& 251.7	$\pm$ 0.3 \\
U-elastic			& Li\`ege cascade				& 223.1	$\pm$ 0.2 \\
U-elastic			& LEP						& 225.4	$\pm$ 0.2 \\
U-elastic			& CHIPS-inelastic				& 250.4	$\pm$ 0.3 \\
\hline
\end{tabular}
\end{center}
\end{table}
		
Based on Table \ref{tab_cpu}, one can observe that the hadronic elastic scattering models 
exhibit similar computational performance, with the exception of the CHIPS model, which is
significantly slower; among the hadronic inelastic models, the  Li\`ege cascade and
the LEP ones are faster than the other options.
											

\section{Conclusion}

A number of epistemic uncertainties have been identified in a survey of Geant4
physics models pertinent to the simulation of proton depth dose, which broadly
represent the variety of approaches to describe proton interactions with matter
in the energy range up to approximately 100 MeV.

In the electromagnetic domain, the epistemic uncertainties affecting the value
of the water mean ionization potential and proton stopping powers derive from
lack of consensus among various theoretical and experimental references
documented in the literature; they generate significant systematic effects on the
longitudinal pattern of energy deposit in the sensitive volume, namely on the
depth of the Bragg peak.

The epistemic uncertainties affecting the hadronic components of the simulation
are related to the intrinsic differences of the modeling approaches and
empirical parameters they contain; the limited validation of the models, and the
unclear distinction between the processes of calibration and validation in the
few published comparisons with experimental data, are the main sources of such
uncertainties.
Their effects on the longitudinal energy deposit are comparable with
experimental uncertainties typical of proton therapy; the largest differences
concern secondary particle spectra.
A significant effect was observed in relation to the mode of nuclear deexcitation; in this
respect, there is a consensus towards modeling it through Fermi break-up for
light nuclei and evaporation for heavier ones.
This approach is implemented in some Monte Carlo codes (e.g. MCNP and FLUKA),
while it is not adopted by default in Geant4; users of this code would benefit from
implementing appropriate settings in their Geant4-based applications to activate
Fermi break-up for the deexcitation of light nuclei, if their simulation use cases
are prone to be affected by the systematic effects highlighted in this study.

The analysis shows how the sensitivity of the simulation to epistemic
uncertainties cannot be determined in absolute terms, rather it depends on the
experimental application environment.
The relatively large differences in the Bragg peak profile associated with the
set of electromagnetic options are practically irrelevant in clinical
practice, which tolerates adjustments of the beam parameters to reproduce a
reference proton range.
However, these differences are relevant to applications where a predictive
role is expected from the simulation, such as Monte Carlo based
treatment planning systems, currently the object of active research, or
radiation protection.
The different secondary particle spectra deriving from the range of available
hadronic options do not affect the main parameter of clinical interest, i.e. the
depth dose distribution, but they are relevant to other aspects of radiation
exposure.

By far the largest effects of physics-related epistemic uncertainties in the
simulation of proton depth dose are observed in relation to modeling
multiple scattering in the beam line.
However, even these effects are relevant only to use cases where the simulation
is invested with predictive role regarding the absolute dose released to the
target; otherwise, common practices, like the normalization of the simulated
dose to a reference value, would mask the epistemic uncertainty associated
with the empirical parameters used to model this process.

The analysis also highlights the importance of a knowledge of the whole
simulation system regarding the effects visible in the sensitive volume.
Interactions in the beam line affect the spectrum of the protons reaching the
sensitive volume and the dose released to it; lack of knowledge of construction
details of the beam line, or epistemic uncertainties in modeling particle
interactions in the passive components of the system, are prone to bias the
simulation outcome.

The results documented in this paper about the different observables
produced by Geant4 physics options identify some experimental requirements 
for the discrimination of their features and their validation.
Experimental measurements of adequate accuracy could reduce the epistemic
uncertainties evidenced in the electromagnetic domain;
relevant data could derive either from a thorough survey of the 
existing literature, or from new, dedicated measurements.
In this respect, it is worthwhile to recall the valuable reference role 
for the validation of electron simulation played
by the high precision measurements of \cite{sandia79} and \cite{sandia80}, which
were originally motivated by the validation of the ITS (Integrated Tiger Series)
\cite{its} Monte Carlo code; similar measurements concerning protons would be
useful to reduce epistemic uncertainties.

The sensitivity analysis documented in the previous sections also provides
guidance to design meaningful test cases for inclusion in the test process
of Monte Carlo systems.
The identification of distributions which expose distinctive
features of the physics models, as well as of others, which are prone to hide
them, is especially useful to designing test cases relevant to monitoring
the effects of changes in some critical parts of the code.


The analysis presented in this paper is a first attempt at estimating
quantitatively the impact of epistemic uncertainties on the considered use
case; further refinements would contribute to better understanding the
problem.
So far, the analysis has considered each source of epistemic uncertainty
individually; nevertheless, it would be worthwhile to evaluate their
combinations, since several systematic contributions could accumulate their
effects to bias the final simulation result.
More refined treatments, e.g. based on the theory of evidence, could shed
additional light on the problem; these methods would be especially useful if
practical constraints hinder the availability of further experimental
measurements to reduce the current uncertainties.

The identification of the epistemic uncertainties embedded in a large-scale
simulation code is far from trivial; design methods facilitating their
identification at early stages of the software development, and their
management in sensitivity analyses, would be beneficial.
To the best of the authors' knowledge, this issue has not been studied yet in
the context of Monte Carlo simulation; techniques like aspect oriented
programming could provide useful paradigms to address it, and 
the inclusion of epistemic uncertainties in the traceability process,
in the context of 
a rigorous software process discipline, would effectively support 
their handling in complex software systems.

Although this paper illustrates the problem of epistemic uncertainties in a
specific simulation use case, the issue it addresses goes beyond the limited
application domain considered in this initial study.
Regarding the simulation of low energy proton interactions, the epistemic
uncertainties discussed in this paper and their effects are likely to affect
other experimental domains as well: from the exposure of electronic components and
astronauts to the space radiation environment, to the problem of radiation
monitoring at particle accelerators.

More generally, the issue of identifying and quantifying epistemic
uncertainties, and their contribution to the overall reliability of simulation
systems, permeates all Monte Carlo application domains.
Monte Carlo simulation - not only for particle transport in detectors, but also
for event generators - is expected to play a critical role in the physics analysis
of LHC data, which involves energies not yet covered by any experimental
measurements in controlled laboratory environments; the development of sound
methods and tools to deal with the epistemic uncertainties embedded in LHC
simulation software appears a major task for the coming years in support of LHC
physics results.

\section*{Acknowledgment}

The authors thank Andreas Pfeiffer for his significant help with data
analysis tools throughout the project; 
Katsuya Amako, Sergio Bertolucci, Luciano Catani, Gloria Corti,
Andrea Dotti, Gunter Folger,
Simone Giani, Vladimir Grichine, Aatos Heikkinen, Alexander Howard, Vladimir
Ivanchenko, Mikhail Kossov, Vicente Lara, Katia Parodi, Alberto Ribon, 
Takashi Sasaki, Vladimir Uzhinsky and Hans-Peter Wellisch for valuable
discussions, and Anita Hollier for proofreading the manuscript.

CERN Library's support  has been essential to this study;
the authors are especially grateful to Tullio Basaglia.

INFN Genova Computing Service
(Alessandro Brunengo, Mirko Corosu, Paolo Lantero and Francesco Saffioti) 
provided helpful technical assistance with the simulation production.

The authors do not intend to express criticism, nor praise regarding 
any of the Monte Carlo codes mentioned in this paper; the purpose of
the paper is limited to documenting technical results.


\end{document}